\documentclass[aps,prd,twocolumn,groupedaddress,nofootinbib,showpacs]{revtex4-1}

\usepackage{graphicx,amssymb}

\usepackage{amsfonts}
\usepackage{amsmath}
\usepackage{color}
\usepackage{euclid}

\begin{document}

\addtolength{\hoffset}{-0.175cm}
\addtolength{\textwidth}{0.35cm}
\title{Minkowski Functionals of Convergence Maps and the Lensing Figure of Merit}

\author{Martina Vicinanza$^{1,2,3,4}$}
\email{mvicinanza@fc.ul.pt}
\author{Vincenzo F. Cardone$^4$}
\email{winnyenodrac@gmail.com}
\author{Roberto Maoli$^{3,5}$}
\email{roberto.maoli@roma1.infn.it}
\author{Roberto Scaramella$^{4,5}$}
\email{roberto.scaramella@oa-roma.inaf.it}
\author{Xinzhong Er$^{6,4}$}
\email{xer@ynu.edu.cn}
\author{Ismael Tereno$^{1,7}$}
\email{iatereno@fc.ul.pt}
\affiliation{$^1$Instituto de Astrofisica e Ci\^{e}ncias do Espa\c{c}o, Universidade de Lisboa, FCUL, Campo Grande, PT1749\,-\,016, Lisbon, Portugal}
\affiliation{$^2$Dipartimento di Fisica, Universit\`{a} di Roma ``Tor Vergata", via della Ricerca Scientifica 1, 00133 - Roma, Italy}
\affiliation{$^3$Dipartimento di Fisica, Universit\`{a} di Roma ``La Sapienza", Piazzale Aldo Moro, 00185 - Roma, Italy}
\affiliation{$^4$I.N.A.F.\,-\,Osservatorio Astronomico di Roma, via Frascati 33, 00040 - Monte Porzio Catone (Roma), Italy}
\affiliation{$^5$I.N.F.N.\,-\,Sezione di Roma 1, Piazzale Aldo Moro, 00185 - Roma, Italy}
\affiliation{$^6$South Western Insitute for Astronomy Research, Yunnan University, Kunming, P.R. China}
\affiliation{$^7$Departamento de F\'{\i}sica, Faculdade de Ci\^encias, Universidade de Lisboa, Edif\'{\i}cio C8, Campo Grande, 1749-016 Lisbon, Portugal}

\begin{abstract}

Minkowski functionals (MFs) quantify the topological properties of a given field probing its departure from Gaussianity. We investigate their use on lensing convergence maps in order to see whether they can provide further insights on the underlying cosmology with respect to the standard second-order statistics, i.e., cosmic shear tomography. To this end, we first present a method to match theoretical predictions with measured MFs taking care of the shape noise, imperfections in the map reconstruction, and inaccurate description of the nonlinearities in the matter power spectrum and bispectrum. We validate this method against simulated maps reconstructed from shear fields generated by the MICE simulation. We then perform a Fisher matrix analysis to forecast the accuracy on cosmological parameters from a joint MFs and shear tomography analysis. It turns out that MFs are indeed helpful to break the $\Omega_{\rm m}$--$\sigma_8$ degeneracy thus generating a sort of chain reaction leading to an overall increase of the Figure of Merit. 

\end{abstract}

\pacs{98.80.-k, 98.80.Es, 95.36.+x}

\maketitle

\section{Introduction}

Weak gravitational lensing (hereafter WL) refers to the small distortions of galaxy images due to the deflection of photons induced by the presence of matter along the line of sight. Such a small distortion can only be detected statistically by averaging over large galaxy samples. In particular, WL induces a scale dependent correlation between galaxies ellipticities that can be quantified through its Fourier transform, i.e., the shear power spectrum. Being dependent on both the growth of structures and the background cosmic expansion, WL has emerged as one of the most promising tools to unveil the mystery of the cause of the cosmic acceleration. The recommendation of the Dark Energy Task Force \cite{Albrecht2006} in favour of WL have increased the interest in WL, with many excellent reviews \cite{Bartelmann2001,Munshi2008,Kilbinger2015} available as a standing point to train the next generation of astronomers and cosmologists in dealing with both the coming data and their interpretation.  After the first pioneering detection in the late 1990s and early 2000 \cite{Brainerd1996,Bacon2000,Kaiser2000,vanwaerbekeetal2000,wittmanetal2000}, WL is now on the verge to enter a new data-driven era. 
 
Recent lensing surveys, such as CFHTLenS \cite{Heymansetal2012} and RCSLenS \cite{Hildebrandtetal2016}, have convincingly demonstrated the technical feasibility of the WL measurements thus paving the way to ongoing larger projects as DES \cite{DES}, KiDS \cite{deJongetal2015}, and HSC \cite{HSC}, which have already released interesting data sets. As fruitful as they are in their own, nowadays surveys can nevertheless be seen as first steps towards next generation ambitious programs both ground- (e.g, LSST \cite{LSST2009}) and space-based (i.e., \Euclid \cite{Laureijs2011} and WFIRST \cite{WFIRST2012}). Next generation programs, with their unprecedented high-quality image, large area and high number density of galaxies are appealing candidates for the next step in WL science. It is indeed now possible to ask whether one can go beyond standard second-order statistics and detect higher-order ones with enough quality to break parameters degeneracies and probe the non-Gaussianity of the lensing convergence field. 

To this end, Minkowski functionals (hereafter MFs) are ideal tools to single out the non-Gaussian features imprinted on the convergence field by the structure formation process. While the power spectrum and, in general, all second-order statistics provide the bulk of the information regarding the underlying cosmology, higher-order statistics, such as the bispectrum and the Fourier transform of any N-point correlation function (with ${\rm N} > 2$), are more commonly used to highlight departures from Gaussianity \cite{Bartoloetal2014}. MFs are determined by the complete set of higher-order terms needed to fully characterize the convergence PDF. In general, MFs describe the morphological features of continuous stochastic fields \cite{Meckeetal1994,Schmalzingetal1997,Schmalzingetal1998}, which, in two dimensions, reduces to the area, the perimeter, and the Euler characteristics of excursion set regions. Closed form analytic expressions exist for the MFs of Gaussian fields \cite{Tomita1986} so that any deviation in measured MFs from these expected values can be taken as a strong evidence of non-Gaussianity of the field of interest. A first remarkable example of the usefulness of these characteristics is provided by the study of the CMB field \cite{Komatsuetal2003,Eriksenetal2004,Hikageetal2008}, while the results of the first attempts in measuring MFs in WL fields have been presented in \cite{MatsubaraJain2001,Satoetal2001,Taruyaetal2002}.

It is worth noting that, at the lowest order, deviations from the Gaussian MFs depend on various one-point moments determined by the bispectrum, while next-to-leading order corrections make higher-order spectra, beyond bispectrum, enter the game with all the technical difficulties and theoretical uncertainties associated with their modelling. Fortunately, even for relatively small angular scales, the leading-order terms are sufficient to satisfactorily model the deviations from Gaussian expectation of the MFs. It is also worth noting that, being topological quantities, their measurement from convergence maps is quite easy, provided the number density of galaxies used to reconstruct the map is large enough. Being so sensitive to higher-order statistics and easy to measure in future surveys, it is worth investigations whether a joint use of MFs and standard shear tomography can strengthen the constraints on the cosmological parameters, in particular on the dark energy equation of state.

The paper is organized as follow. In Sect.\,II we give an overview on theoretical MFs formulae. In Sect.\,III we develop our approach of deriving phenomenological relations between theoretical MFs and observed ones, taking into account noise and map reconstruction bias. Using the simulated maps and mock data presented in Sect.\,IV, we calibrate and validate the procedure in Sect.\.V. We investigate the cosmological constraining power of the MFs through a Fisher matrix based forecast analysis in Sect.\,VI, and we conclude in Sect.\,VII.

\section{Minkowski functionals}

Given an N-dimensional Euclidean space, let us consider the set of convex bodies embedded in it and the so-called convex ring ${\cal{R}}$ formed by all finite unions of convex bodies. Any polyconvex ${\cal{B}}$ on the convex ring may be characterized by a scalar functional ${\cal{M}}$ which is motion invariant (i.e., its value does not depend on the position and orientation in the space), additive (i.e., the functionals of the union of two bodies are the sum of the single values), and conditionally continuous (i.e., the functional of convex approximations of a convex body converges to the functional of the body itself). Hadwiger \cite{hadwiger2013vorlesungen} demonstrated that any functional with these properties may be written as a linear combination of only N + 1 functionals that satisfy these properties. 

This basis is commonly represented by the MFs since they have a quite convenient geometrical interpretation for $1 \le {\rm N} \le 3$. When considering a smooth two-dimensional field $u(\vec{x})$, with ${\bf x}$ the position on the 2D plane or the 2D sphere, the MFs account for the geometry and the topology of the excursion set $Q_\nu$ of the field formed by the region where the field $u(\vec{x})$ is larger than a threshold value $\nu$. We can then define the Euler characteristic $\chi$ as the line integral of curvature of the border $\partial Q_\nu$. The MFs are given by
\begin{displaymath}
V_0(Q_\nu) = \int_{Q_\nu}{\diff a} \;  ,
\end{displaymath}
\begin{displaymath}
V_1(Q_\nu) = \frac{1}{4} \int_{\partial Q_\nu}{\diff l} \; ,
\end{displaymath}
\begin{displaymath}
V_2(Q_\nu) = \frac{1}{2 \pi} \int_{\partial Q_\nu}{{\cal{K}} \; \diff l} \; ,
\end{displaymath}
where $\diff a$ and $\diff l$ are the surface and the line element of the excursion set along $\partial Q_\nu$, and ${\cal{K}}$ is the local geodesic curvature of $\partial Q_\nu$. Geometrically, the three MFs may be interpreted as the area, the perimeter, and the Euler characteristics (i.e., the number of connected regions minus the number of holes) of the excursion set regions. 

In order to theoretically estimate MFs, let us first start by the simplest case of a Gaussian field, which has been extensively studied in the literature \cite{adler1981,Tomita1986}. In two dimensions, the MFs for the excursion set with threshold $\nu$ are given by \cite{Tomita1986}
\begin{eqnarray}
V_{n}^{\rm G}(\nu)  & = & \frac{1}{(2 \pi)^{(n + 1)/2}} \, \frac{\omega_2}{\omega_{2 - n} \omega_n} \, 
\left ( \frac{\sigma_1}{\sqrt{2} \sigma_0} \right )^{n} \nonumber \\
 & \times & \exp{\left ( - \frac{\nu^2}{2} \right )} \, {\cal{H}}_{n-1}(\nu)
\label{eq: GaussMFs}
\end{eqnarray}
with $\omega_n=\pi^{n/2}\left[\Gamma(n/2+1)\right]^{-1}$ so that it is $\omega_0 = 1$, $\omega_1 = 2$, $\omega_2 = \pi$. This expression is valid for a field $u(\vec{x})$ with null mean, and $(\sigma_0, \sigma_1)$ denote the variance of the field and its covariant derivative, respectively. ${\rm{\cal{H}}}_n(\nu)$ are Hermite polynomials defined through the following recurrence relations\,:
\begin{displaymath}
{\cal{H}}_{-1}(\nu) = \sqrt{\frac{\pi}{2}}\, \exp{\left ( \frac{\nu^2}{2} \right )}\, {\rm erfc}{\left ( \frac{\nu}{\sqrt{2}} \right )} \; ,
\end{displaymath}
\begin{displaymath}
{\cal{H}}_n(\nu) = (-1)^{n}\, \exp{\left ( \frac{\nu^2}{2} \right )}\, \frac{\diff ^n}{\diff \nu^n}\, \left [ \exp{\left ( - \frac{\nu^2}{2} \right )} \right ] \ .
\end{displaymath}
The  fields one is interested in are not Gaussian and Eq.(\ref{eq: GaussMFs}) incorrectly match the observed values. However, if the field is only mildly non-Gaussian, a perturbative expansion can be used. Considering hereafter the convergence field $\kappa$, the MFs can be written as the sum of two terms, thus reading
\begin{equation}
V_{n}(\nu) = V_{n}^{\rm G}(\nu) + \delta V_{n}(\nu)\; .
\label{eq: mfsum}
\end{equation}
The deviation from the Gaussian prediction can be expanded in terms of\footnote{Since the convergence field has zero mean, its variance is simply the expectation value of $\kappa^2$.} $\sigma_0 = \langle \kappa^2 \rangle$ as 
\begin{eqnarray}
\delta V_{n}(\nu) & = & 
\frac{1}{(2 \pi)^{(n + 1)/2}} \, \frac{\omega_2}{\omega_{2 - n} \omega_n} \, 
\left ( \frac{\sigma_1}{\sqrt{2} \sigma_0} \right )^{n} \,\exp{\left ( - \frac{\nu^2}{2} \right )} \nonumber \\
 & \times & \left [ \delta V_n^{(2)}(\nu) \sigma_0 + \delta V_{n}^{(3)}(\nu) \sigma_0^2 + \ldots \right ]
\label{eq: deltavn}
\end{eqnarray}
with $\sigma_{1}^{2} = \left \langle (\nabla \kappa)^2 \right \rangle$. To the lowest order in $\sigma_0$, the coefficient of the correction term reads
\begin{eqnarray}
\delta V_{n}^{(2)}(\nu) & = &
S^{(0)} \, {\cal{H}}_{n + 2}(\nu)/6 \nonumber \\
 & + &  n\, S^{(1)}\, {\cal{H}}_{n}(\nu)/3  \nonumber \\
 & + & n\, (n - 1)\, S^{(2)}\, {\cal{H}}_{n - 2}(\nu)/6 \; ,
\label{eq: deltaV2}
\end{eqnarray}
where $S^{(n)}$ are generalized skewness quantities defined from of the convergence field and its derivatives
\begin{eqnarray}
S^{(0)} & = & \frac{\langle \kappa^3 \rangle}{\sigma_{0}^{4}} \; , \\
S^{(1)} & = & -\frac{3}{4}\, \frac{\langle \kappa^2 \nabla^2 \kappa \rangle}{\sigma_{0}^{2} \sigma_{1}^{2}} \; , \\
S^{(2)} & = & - 3 \, \frac{\langle (\nabla \kappa) \cdot (\nabla \kappa) ( \nabla^2 \kappa) \rangle}{\sigma_{1}^{4}} \; .
\label{eq: skewdef}
\end{eqnarray}
Both the variance terms $\sigma_n$ and the the generalized skewness parameters $S^{(n)}$ can be expressed in terms of the polyspectra of the field. For the variances, it is indeed \cite{Munshi12}
\begin{equation}
\sigma_n^2 = \frac{1}{4 \pi}\, \sum_{\ell}{(2 \ell + 1) [\ell (\ell + 1)]^n\, {\cal{C}}(\ell)\, {\cal{W}}^{2}(\ell)}
\label{eq: variance}
\end{equation}
where ${\cal{C}}(\ell)$ is the lensing convergence power spectrum for sources at redshift $z_{\rm s}$, and ${\cal{W}}(\ell)$ is the Fourier transform of the smoothing filter. The cosmological information is contained in ${\cal{C}}(\ell)$, which is given by
\begin{equation}
{\cal{C}}(\ell) = \frac{c}{H_0} \int_{0}^{z_{\rm s}}
{\frac{W^2(z)}{r^2(z) E(z)}\, P_{\rm NL}\left [ \frac{\ell}{\chi(z)}, z \right ]\, \diff z}
\label{eq: defconvps}
\end{equation}
with 
%{\rm M}
\begin{equation}
W(z) = \frac{3}{2} \, \Omega_{\rm m} \left ( \frac{H_0}{c} \right )^2  \chi(z) \left [1 - \frac{\chi(z)}{\chi(z_{\rm s})} \right ]\; ,
\label{eq: defkernel}
\end{equation}
where $E(z) = H(z)/H_0$ is the dimensionless Hubble function, $\chi(z)$ the comoving distance, $r(z)$ the comoving angular diameter distance, and $P_{\rm NL}(k, z)$ the nonlinear matter power spectrum evaluated in $k = \ell/\chi(z)$ because of the Limber approximation. Note that hereafter we will assume a spatially flat universe. We will use a Gaussian filter to smooth the map, i.e., 
\begin{equation}
{\cal{W}}(\ell) = \exp{\left ( - \ell^2 \sigma_{\rm s}^2 \right )}
\label{eq: wlgauss}
\end{equation}
with $\sigma_{\rm s}$ the smoothing length (typically given in arcmin).

Since variances are related to second-order moments and hence to 2-point statistics, it is not surprising that generalized skewness quantities (which are connected with third-order moments) can be expressed as a function of the Fourier transform of the 3-point correlation function. Indeed, it is 
\begin{equation}
S^{(n)} = \sum_{\ell}{(2 \ell + 1)\, {\cal{S}}^{(n)}(\ell)}
\label{eq: sncalc}
\end{equation}
where, adopting a compact notation, we get
\begin{equation}
{\cal{S}}^{(n)}(\ell) =  
\sum_{\ell_1, \ell_2}{
\frac{s_n (\ell, \ell_1, \ell_2)\, {\cal{B}}(\ell, \ell_1, \ell_2)\, \tilde{{\cal{W}}}(\ell, \ell_1, \ell_2)}
{{\cal{N}}_{n}(\sigma_0, \sigma_1)}} \; ,
\label{eq: snelle}
\end{equation}
with 
\begin{equation}
{\cal{N}}_{n}(\sigma_0, \sigma_1) = \left \{ 
\begin{array}{ll}
12 \pi \sigma_{0}^{4} & \ \ n = 0 \\ 
 & \\
16 \pi \sigma_{0}^{2} \sigma_{1}^{2} & \ \ n = 1 \\
 & \\
8 \pi \sigma_{1}^{4} & \ \ n = 2 \\
\end{array}
\right . \ ,
\label{eq: normcst}
\end{equation}
and 
\begin{eqnarray}
s_0 & = & 1  \nonumber \\
 & \nonumber \\ 
s_1 & = & \ell\, (\ell + 1) + \ell_1\, (\ell_1 + 1) + \ell_2\, (\ell_2 + 1) \nonumber \\
 & \nonumber \\
s_2 & = & [\ell\, (\ell + 1) + \ell_1 \, (\ell_1 + 1) - \ell_2\, (\ell_2 + 1)]\, \ell_2\, (\ell_2 + 1) + {\rm cp} \nonumber 
\end{eqnarray}
where "cp" denotes cyclic permutation. In Eq.(\ref{eq: snelle}), the cosmological information is coded into the convergence bispectrum given by
\begin{eqnarray}
{\cal{B}}(\ell_1, \ell_2, \ell_3) & = & \frac{c}{H_0}  \\ 
 & \times & \int_{0}^{z_s}
{\frac{W^3(z)}{r^4(z) E(z)}\, B_{\rm NL}\left [ \frac{\ell_1}{\chi(z)}, \frac{\ell_2}{\chi(z)}, \frac{\ell_3}{\chi(z)} \right ]\; \diff z} \nonumber 
\label{eq: convbps}
\end{eqnarray}
with $B_{\rm NL}(k_1, k_2, k_3, z)$, the matter bispectrum, evaluated at $k_i = \ell_i/\chi(z)$ because of the Limber approximation. The contribution of each multipole to the sum in Eq.(\ref{eq: snelle}) is weighted by the function
\begin{equation}
{\tilde{W}}(\ell_1, \ell_2, \ell_3) = {\cal{J}}(\ell_1, \ell_2, \ell_3)\, {\cal{W}}(\ell_1) {\cal{W}}\,(\ell_2)\, {\cal{W}}(\ell_3)
\label{eq: defwtilde}
\end{equation}
with
\begin{equation}
{\cal{J}}(\ell_1, \ell_2, \ell_3) = \frac{{\cal{I}}^2(\ell_1, \ell_2, \ell_3)}{2 \ell_3 + 1} \; ,
\label{eq: defjelle}
\end{equation}
and
\begin{equation}
{\cal{I}} = \sqrt{\frac{(2 \ell_1 + 1)\,(2 \ell_2 + 1)\,(2 \ell_3 + 1)}{4 \pi}}
\left (
\begin{array}{lll}
\ell_1 & \ell_2 & \ell_3 \\
 & & \\
0 & 0 & 0 \\
\end{array}
\right ) \; .
\label{eq: defielle}
\end{equation}
The second term in the right-hand side of Eq.(\ref{eq: defielle}) denotes the Wigner-3j symbols. These are introduced to take into account the fact that the bispectrum is only not null when the three points correlated define a triangle so that $\vec{k}_1 + \vec{k}_2 + \vec{k}_3 = 0$.

The next-to-leading order corrections $\delta V_{n}^{(3)}$ in Eq.(\ref{eq: deltavn}) are determined by four generalized kurtosis parameters, which can be constructed from the trispectrum. Although explicit formulae are available \cite{Munshi12}, we do not consider these terms and cut the expansion in Eq.(\ref{eq: deltavn}) to linear order in $\sigma_0$. Such a choice is motivated by our underlying philosophy aiming at both simplifying the MFs estimate, and reducing the uncertainties on the nonlinear corrections. As it will be investigated in detail later, we will look for linear relations among the theoretical MFs thus computed and the measured one from simulated maps. Should such a relation indeed be found out, we can then rely on the linear order approximation to estimate the MFs for a given set of cosmological parameters.

\section{Matching theory and observations}

Eqs.(\ref{eq: mfsum})--(\ref{eq: deltaV2}) allow us to evaluate the theoretical MFs for a given set of cosmological parameters provided one has a recipe for the matter power spectrum and bispectrum. Actually, there are a number of reasons why the theoretical MFs thus computed do not match the ones measured from real maps. We detail some of them below. 

\begin{enumerate}

\item[i.]{{\it Nonlinearities in matter power spectrum and bispectrum.} While modelling the matter power spectrum in the linear regime is an easy task, the impact of nonlinearities is far to be considered under full control. Although well tested recipes do exist \cite{Takahashi2012,Mead15}, they have been tested for the matter power spectrum, while much less work has been done for the bispectrum. As a consequence, we should allow for the possibility that the actual $P_{\rm NL}(k, z)$ and $B_{\rm NL}(k_1, k_2, k_3, z)$ we are using to predict the convergence polyspectra (and hence the variances and the generalized skewness parameters) are different from our theoretical model for the nonlinear corrections.}

\item[ii.]{{\it Finiteness of the triangle configurations.} The sum in Eq.(\ref{eq: sncalc}) formally extends over all the combinations of $(\ell_1, \ell_2, \ell_3)$ defining a triangle in the $(\vec{k}_1, \vec{k}_2, \vec{k}_3)$ space. Actually, for computing time reasons, only a finite number of combinations is included in the sum so that the estimated $S^{(n)}$ is smaller than the actual one. It is worth noting, however, that the different terms in the sum are weighted through the smoothing function, which cuts out the contributions at high $\ell$ so that we expect the mismatch to be small.}

\item[iii.]{{\it Convergence of the series expansion.} Eq.(\ref{eq: deltavn}) shows that the deviations from the Gaussian case can be expanded in series of $\sigma_0$. It is clear that the higher the order of the expansion, the better the MFs from Eq.(\ref{eq: mfsum}) approximate the true (unknown) ones. However, there are both practical and theoretical motivations to halt the series expansion to the first order in $\sigma_0$. Indeed, the higher is the order of the expansion, the higher is the order of the polyspectra one has to compute. Going to second order in $\sigma_0$ asks for the trispectrum, which is hard to model. No models at all exist for $n > 3$ n-point correlation functions so that one should resort to N-body simulations to get a hint. Moreover,  even for the trispectrum case, nonlinear corrections are largely unknown so that one does not know how to deal with this problem (not to mention the large computation time needed to estimate the trispectrum in a fitting pipeline). We have therefore stopped our expansion at the first order and looked for a way of correcting a posteriori the missing terms.}

\item[iv.]{{\it Map making from noisy data.} Convergence maps are not directly available, but must be reconstructed from the shear data using a suitable algorithm such as the popular KS we have used here. Even under ideal conditions, the map making procedure leads to a biased reconstruction of the convergence field so that the measured MFs will not be the same as the theoretical ones. Moreover, ideal conditions are never met in practice because of unavoidable noise in the data which introduces a further bias which can not be quantified a priori.}

\end{enumerate}

Motivated by these considerations, we develop below a semianalytical approach to predict the observed MFs from those evaluated using Eqs.(\ref{eq: mfsum})--(\ref{eq: deltavn}) with the first order correction estimated from Eqs.(\ref{eq: skewdef})--(\ref{eq: defielle}). This approach introduces some calibration parameters that we will determine by comparing with MFs measured on simulated maps. The impatient reader can skip the remaining of the section, and move to Sect.\,V where we validate the final result against simulated data.

\subsection{MFs for a noisy convergence field}

As a starting point, we assume that the observed convergence $\kappa_{\rm obs}$ is related to the actual one $\kappa$ by a simple linear relation
\begin{equation}
\kappa_{\rm obs} = (1 + m_{\kappa})\, \kappa + {\cal{N}}
\label{eq: kappaobs}
\end{equation}
where ${\cal{N}}$ is a noise term with null mean value, and we have removed the zeropoint of the $\kappa_{\rm obs}$ vs $\kappa$ relation in the no noise case so that both $\kappa$ and $\kappa_{\rm obs}$ average to zero\footnote{Note that this is expected for the actual convergence because of isotropy of the Universe. This can also not be true for the observed convergence field $\kappa_{\rm obs}$ , but one can always redefine the field by subtracting the mean value.}. 

Eq.(\ref{eq: kappaobs}) naively leads to the following relations for the derivatives of interest 
\begin{displaymath}
\nabla \kappa_{\rm obs} =(1 + m_{\kappa})\,  \nabla \kappa + \nabla {\cal{N}} \; ,
\end{displaymath}
\begin{displaymath}
\nabla^2 \kappa_{\rm obs} =(1 + m_{\kappa}) \, \nabla^2 \kappa + \nabla^2 {\cal{N}} \; ,
\end{displaymath}
which will be useful later. Note that, since the noise has not expected to have a well defined pattern, the terms $\nabla {\cal{N}}$ and $\nabla^2 {\cal{N}}$ can take quite large values. 

The next step is to estimate the variance of the observed field and of its derivative. Assuming the signal and the noise are not correlated with each other\footnote{Under this assumption, we can put to zero the expectation value of all the $\kappa^a\, {\cal{N}}^b$-like terms.}, we get 
\begin{equation}
\sigma_{0,{\rm obs}}^{2} = \langle \kappa_{\rm obs}^2 \rangle = (1 + m)^2\, \sigma_{0}^{2} + \langle {\cal{N}}^2 \rangle \; ,
\label{eq: sigmazeroobs}
\end{equation}
\begin{equation}
\sigma_{1,\rm obs}^{2} = \langle (\nabla \kappa_{\rm obs})^2 \rangle = (1 + m)^2\, \sigma_{1}^{2} + \langle (\nabla {\cal{N}})^2 \rangle \; ,
\label{eq: sigmaunoobs}
\end{equation}
where in Eq.(\ref{eq: sigmaunoobs}) we have assumed that also the field derivatives are uncorrelated. We can now go on computing the generalized skewness parameters $S^{(n)}_{\rm obs}$ of the observed convergence field. Let us start with $S^{(0)}_{\rm obs}$ proceeding as follows
\begin{eqnarray}
S^{(0)}_{\rm obs} & = & \frac{\langle \kappa_{\rm obs}^3 \rangle}{\sigma_{0,{\rm obs}}^{4}} \nonumber \\
 & = & \frac{(1 + m_{\kappa})^3\, \langle \kappa^3 \rangle + \langle {\cal{N}}^3 \rangle}
{[(1 + m_{\kappa})^2\, \sigma_{0}^{2} + \langle {\cal{N}}^2 \rangle]^2} \nonumber \\
 & = & \frac{[(1 + m_{\kappa})^3\, \langle \kappa^3 \rangle + \langle {\cal{N}}^3 \rangle]\,/\,\sigma_{0}^{4}}
{[(1 + m_{\kappa})^2\, \sigma_{0}^{2} + \langle {\cal{N}}^2 \rangle]^2\,/\,\sigma_{0}^{4}} \nonumber \\
 & = & \frac{(1 + m_{\kappa})^3\, S^{(0)} + S^{(0)}_{N}\, (\sigma_{0{\rm N}}/\sigma_0)^4}{[(1 + m_{\kappa})^2 + (\sigma_{0{\rm N}}/\sigma_0)^2]^2}
\label{eq: s0obs}
\end{eqnarray}
where $\sigma_{0{\rm N}} = \langle {\cal{N}}^2 \rangle$ and $S^{(0)}_{\rm N} = \langle {\cal{N}}^3 \rangle/\sigma_{0{\rm N}}^{4}$ are the variance and zeroth order generalized skewness of the noise field, respectively.

To move to the next skewness quantity, we first need to evaluate the expectation value of $\kappa_{\rm obs}^2\, \nabla^2 \kappa_{\rm obs}$. It turns out to be
\begin{eqnarray}
\langle \kappa_{\rm obs}^2\, \nabla^2 \kappa_{\rm obs} \rangle & = & 
(1 + m_{\kappa})^3\, \langle \kappa^2 \, \nabla^2 \kappa \rangle  
+ (1 + m_{\kappa})^2\, \langle \kappa^2 \rangle \,\langle \nabla^2 {\cal{N}} \rangle \nonumber \\
 & + & (1 + m_{\kappa})\, \langle \nabla^2 \kappa \rangle \,\langle {\cal{N}}^2 \rangle 
+ \langle {\cal{N}}^2\, \nabla^2{\cal{N}} \rangle \; ,
\label{eq: k2nk2obs}
\end{eqnarray}
where we have used the property that all the cross terms $\kappa\, \nabla {\cal{N}}$ and ${\cal{N}}\, \nabla \kappa$ have zero expectation value since the convergence and noise fields are uncorrelated. Inserting Eq.(\ref{eq: k2nk2obs}) into the definition of $S^{(1)}$, and using the previously obtained expressions for the variances $(\sigma_{0,{\rm obs}},\, \sigma_{1,{\rm obs}})$, we find after some lengthy algebra,  
\begin{eqnarray}
S^{(1)}_{\rm obs} & = & \left \{ (1 + m_{\kappa})^3\, S^{(1)} 
+ S^{(1)}_{N}\, \left ( \frac{\sigma_{0{\rm N}}}{\sigma_0} \right )^2\, \left ( \frac{\sigma_{1{\rm N}}}{\sigma_{1}} \right )^2 \right . \nonumber \\
 & - & \left . \frac{3}{4}\, \left [ \frac{(1 + m_{\kappa})\, \langle \nabla^2 {\cal{N}} \rangle}{\sigma_{1}^{2}}  
+ \frac{\langle \nabla^2 \kappa \rangle}{\sigma_{1}^{2}} \left ( \frac{\sigma_{0{\rm N}}}{\sigma_{0}} \right )^2 \right ] \right \} \nonumber \\
 & \times & \left [ (1 + m_{\kappa})^2 + \left (\frac{\sigma_{0{\rm N}}}{\sigma_0} \right )^2 \right ]^{-1} \nonumber \\ 
 & \times &  \left [ (1 + m_{\kappa})^2 + \left (\frac{\sigma_{1{\rm N}}}{\sigma_1} \right )^2 \right ]^{-1} \nonumber  \; .
\end{eqnarray}
This can be recast in a more illuminating way noting that $\langle \nabla^2 \kappa \rangle = \sigma_{2}^{2}$, i.e., it is the variance of the Laplacian of the convergence field. Doing the same for the noise field, we finally get
\begin{eqnarray}
S^{(1)}_{\rm obs} & = & \left \{ (1 + m_{\kappa})^3\, S^{(1)} 
+ S^{(1)}_{N}\, \left ( \frac{\sigma_{0{\rm N}}}{\sigma_0} \right )^2 \left ( \frac{\sigma_{1{\rm N}}}{\sigma_{1}} \right )^2 \right . \nonumber \\
 & - & \left . \frac{3}{4}\, \left ( \frac{\sigma_{2}}{\sigma_{1}} \right )^2 
\left [ (1 + m_{\kappa}) \left ( \frac{\sigma_{2{\rm N}}}{\sigma_2} \right )^2 
+  \left ( \frac{\sigma_{0{\rm N}}}{\sigma_0} \right )^2 \right ] \right \} \nonumber \\
 & \times & \left [ (1 + m_{\kappa})^2 + \left (\frac{\sigma_{0{\rm N}}}{\sigma_0} \right )^2 \right ]^{-1} \nonumber \\ 
 & \times &  \left [ (1 + m_{\kappa})^2 + \left (\frac{\sigma_{1{\rm N}}}{\sigma_1} \right )^2 \right ]^{-1} \nonumber  \; ,
\label{eq: s1obs}
\end{eqnarray}
which shows that the first order generalized skewness gets contributions from higher-order variances when noise is present. The same method also allows us to compute the second-order skewness. Skipping the detailed derivation we arrive at
\begin{eqnarray}
S^{(2)}_{\rm obs} & = & \left \{ (1 + m_{\kappa})^3\, S^{(2)} + S^{(2)}_{N}\, \left ( \frac{\sigma_{1{\rm N}}}{\sigma_{1}} \right )^4 
- 3\,(1 + m_{\kappa})  \right . \nonumber \\
 & \times & \left ( \frac{\sigma_2}{\sigma_1} \right )^2  \left .
\left [ (1 + m_{\kappa})^2 \,\left ( \frac{\sigma_{2{\rm N}}}{\sigma_2} \right )^2 + \left ( \frac{\sigma_{1{\rm N}}}{\sigma_1} \right )^2 \right ] \right \} \nonumber \\
 & \times &  \left [ (1 + m_{\kappa})^2 + \left (\frac{\sigma_{1{\rm N}}}{\sigma_1} \right )^2 \right ]^{-2} \nonumber  \; ,
\label{eq: s2obs}
\end{eqnarray}
which again shows the mixing of variances of different orders caused by the presence of the noise. Note that setting the noise quantities to zero reduces the above formulae to $S^{(n)}_{\rm obs} = (1 + m_{\kappa})^3\, S^{(n)}$ as expected. 

We can now obtain the MFs for the observed convergence field plugging Eqs.(\ref{eq: sigmazeroobs}) and (\ref{eq: sigmaunoobs}) for the variances $(\sigma_{0,{\rm obs}},\, \sigma_{1,{\rm obs}})$, and (\ref{eq: s0obs})-(\ref{eq: s2obs}) for the generalized skewness $S^{(n)}_{\rm obs}$ into Eqs.(\ref{eq: mfsum})--(\ref{eq: deltaV2}) for the perturbative MFs. Collecting the relevant terms, we finally get 
\begin{eqnarray}
V_{0,{\rm obs}}(\nu) & = & \frac{1}{\sqrt{2 \pi}}\, \exp{\left ( - \frac{\nu^2}{2} \right )} \nonumber \\
 & \times & \Bigg \{ {\cal{H}}_{-1}(\nu) + 
\left [ (1 + m_{\kappa})^2 + {\cal{R}}_{0}^{2} \right ]^{1/2}\, \sigma_0  \nonumber \\
 & \times &  \frac{[(1 + m_{\kappa})^3\, S^{(0)} + \tilde{S}^{(0)}]\, {\cal{H}}_2(\nu)}
{6\, [(1 + m_{\kappa})^2 + {\cal{R}}_{0}^{2}]^2} \Bigg \} \; , 
\label{eq: v0obs}
\end{eqnarray}
\begin{eqnarray}
V_{1,{\rm obs}}(\nu) & = & \frac{1}{8}\, \left ( \frac{\sigma_1}{\sqrt{2} \sigma_0} \right )\, \exp{\left ( - \frac{\nu^2}{2} \right )} \nonumber \\
 & \times & \left [ \frac{(1 + m_{\kappa})^2 + {\cal{R}}_{1}^{2}}{(1 + m_{\kappa})^2 + {\cal{R}}_{0}^{2}} 
\right ]^{1/2} \nonumber \\
 & \times & \Bigg \{ {\cal{H}}_{0}(\nu) + \left [ (1 + m_{\kappa})^2 + {\cal{R}}_{0}^{2} \right ]^{1/2} 
\sigma_0  \nonumber \\
 & \times & \left [ \frac{[(1 + m_{\kappa})^3\, S^{(0)} + \tilde{S}^{(0)}]\, {\cal{H}}_3(\nu)}
{6\, [(1 + m_{\kappa})^2 + {\cal{R}}_{0}^{2}]^2} \right . \nonumber \\
 & + & \left . \frac{[(1 + m_{\kappa})^3\, S^{(1)} + \tilde{S}^{(1)}]\, {\cal{H}}_{1}(\nu)}
{3\, [(1 + m_{\kappa})^2 + {\cal{R}}_{0}^{2}]\, 
[(1 + m_{\kappa})^2 + {\cal{R}}_{1}^{2}]} \right ] \Bigg \} \ , 
\label{eq: v1obs}
\end{eqnarray}
\begin{eqnarray}
V_{2,{\rm obs}}(\nu) & = & \frac{1}{(2 \pi)^{3/2}}\, \left ( \frac{\sigma_1}{\sqrt{2} \sigma_0} \right )^2\, \exp{\left ( - \frac{\nu^2}{2} \right )} \nonumber \\
 & \times & \left [ \frac{(1 + m_{\kappa})^2 + {\cal{R}}_{1}^{2}}{(1 + m_{\kappa})^2 + {\cal{R}}_{0}^{2}} 
\right ] \nonumber \\
 & \times & \Bigg \{ {\cal{H}}_{1}(\nu) + \left [ (1 + m_{\kappa})^2 + {\cal{R}}_{0}^{2} \right ]^{1/2} \,
\sigma_0  \nonumber \\
 & \times & \left [ \frac{[(1 + m_{\kappa})^3\, S^{(0)} + \tilde{S}^{(0)}]\, {\cal{H}}_4(\nu)}
{6\, [(1 + m_{\kappa})^2 + {\cal{R}}_{0}^{2}]^2} \right . \nonumber \\
 & + & \frac{2\, [(1 + m_{\kappa})^3\, S^{(1)} + \tilde{S}^{(1)}]\, {\cal{H}}_{2}(\nu)}
{3\, [(1 + m_{\kappa})^2 + {\cal{R}}_{0}^{2}]\, [(1 + m_{\kappa})^2 + {\cal{R}}_{1}^{2}]}  \nonumber \\
 & + & \left . \frac{[(1 + m_{\kappa})^3\, S^{(2)} + \tilde{S}^{(2)}]\, {\cal{H}}_0(\nu)}
{3\, [(1 + m_{\kappa})^2 + {\cal{R}}_{1}^{2}]^2} \right ] \Bigg \} \; , 
 \label{eq: v2obs}
\end{eqnarray}
where, to shorten the notation, we have introduced the variance ratios ${\cal{R}}_i = \sigma_{i {\rm N}}/\sigma_i$, and defined the tilde skewness parameters as
\begin{equation}
\tilde{S}^{(0)} = S^{(0)}_{\rm N}\, {\cal{R}}_{0}^{4} \; ,
\label{eq: s0tilde}
\end{equation}
\begin{equation}
\tilde{S}^{(1)} = S^{(1)}_{\rm N}\, {\cal{R}}_{0}^{2}\, {\cal{R}}_{1}^{2} 
- (3/4)\, \sigma_{21}^{2}\, \left [(1 + m_{\kappa})\, {\cal{R}}_{2}^{2}  + {\cal{R}}_{0}^{2} \right ] \; ,
\label{eq: s1tilde}
\end{equation}
\begin{equation}
\tilde{S}^{(2)} = S^{(2)}_{\rm N}\, {\cal{R}}_{1}^{4} 
- 3\, (1 + m_{\kappa})\, \sigma_{21}^{2}\, \left [(1 + m_{\kappa})\, {\cal{R}}_{2}^{2}  + {\cal{R}}_{1}^{2} \right ] \; ,
\label{eq: s2tilde} 
\end{equation}
with $\sigma_{21} = \sigma_{2}/\sigma_{1}$. Eqs.(\ref{eq: v0obs})--(\ref{eq: s2tilde}) are the main results of this section. They make it possible to estimate the MFs of the observed convergence field in terms of the statistical properties, namely variances $\sigma_{n}$ and generalized skewness parameters $S^{(n)}$ (with $n = 0, 1, 2$) of both the actual convergence field and the noise. 

\subsection{Theoretical quantities and nuisance parameters}

The statistical properties of the convergence field can be evaluated as described in Sect.\,II for any given cosmological model. There are, however, some subtle issues that make this derivation far from trivial. 

First, we note that the $\sigma_{n}$ variances depend through Eq.(\ref{eq: variance}) on the convergence power spectrum ${\cal{C}}(\ell)$ given by Eq.(\ref{eq: defconvps}). As a consequence, one needs to know how to compute the matter power spectrum $P_{\rm NL}(k, z)$. While this is quite easy in the linear regime, the integration actually goes deep into the nonlinear regime so that one has to choose a recipe to model nonlinearities. We here follow the revised HaloFit method \cite{Takahashi2012}, but warn the reader that this has been actually tested only up to $k \le 1 h/{\rm Mpc}$, while we use it far beyond this limit. Moreover, even in this regime, different choices are possible with the Halo Model inspired formalism of Mead et al. \cite{Mead15} being the most recent example. We can qualitatively explore the impact of this uncertainty assuming that the true (unknown) power spectrum $P_{\rm NL}(k ,z)$ is related to $P_{\rm HF}(k, z)$, which we compute with HaloFit, as 
\begin{displaymath}
P_{\rm NL}(k, z) = (1 + m_{\rm HF})\, P_{\rm HF}(k, z) + \pi_{\rm HF}
\end{displaymath}
where $(m_{\rm HF}, \pi_{\rm HF})$ are constant quantities. Such a linear relation translates the qualitative consideration that the HaloFit power spectrum is a good approximation of the one measured from N-body simulations. We then have
\begin{displaymath}
\sigma_{n}^{2} = (1 + m_{\rm HF})\, \sigma_{n,{\rm HF}}^{2} + \sigma_{\pi}^{2}
\end{displaymath}
with $\sigma_{\pi}$ a constant offset. Unless $\pi_{\rm HF}$ is large, we can safely assume that the product $\pi_{\rm HF}\, {\cal{W}}^2(\ell)$ in Eq.(\ref{eq: variance}) quickly fades away so that $\sigma_{\pi}$ can be neglected. As a consequence, the ratios $\sigma_1/\sigma_0$ and $\sigma_{2}/\sigma_1$ entering Eqs.(\ref{eq: v0obs})--(\ref{eq: v2obs}) are unbiased, while the constant factor $(1 + m_{\rm HF})$ can be absorbed into a redefinition of $m_{\kappa}$. We will, therefore, simply assume that $(m_{\rm HF}, \pi_{\rm HF}) = (0, 0)$, i.e., that our theoretical model is capable of giving unbiased estimates of the convergence field variances. 

A similar argument also holds for the theoretical generalized skewness parameters $S^{(n)}$ which depend on the matter bispectrum $B_{\rm NL}(k_1, k_2, k_3)$. Apart from the uncertainties in modelling this quantity in the nonlinear regime, we now have the further complication that the sum in Eq.(\ref{eq: sncalc}) should include all $(\ell_1, \ell_2, \ell_3)$ triangular combinations which are potentially an infinite number. As a first approximation, we will assume that 
\begin{displaymath}
S^{(n)} = (1 + s_n)\, S^{(n)}_{\rm th} + t_n
\end{displaymath}
where $S^{(n)}_{\rm th}$ is the value estimated for the fiducial bispectrum model and finite number of triangular combinations, and $(s_n, t_n)$ are constants. Inserting this relation into Eqs.(\ref{eq: v0obs})--(\ref{eq: v2obs}) does not alter their structure, but only change the multiplicative and additive terms as follows
\begin{displaymath}
(1 + m_{\kappa})^3\, S^{(n)} \rightarrow (1 + m_{\kappa})^3\, (1 + s_n)\, S^{(n)}_{\rm th} \ \ ,  \ \ 
 \tilde{S}^{(n)} \rightarrow \tilde{S}^{(n)} + t_n \ \ .
\end{displaymath}
While this is a quite straightforward modification, we nevertheless prefer to still set $s_n = t_n = 0$. Such a choice reduces the number of unknown parameters, but it is also consistent with our underlying philosophy. We indeed want to see whether, given the theoretical armory at our disposal, we can match theoretical and measured MFs through a set of equations specified by a number of nuisance parameters as small as possible. 

We have then to model the statistical properties of the noise field. Needless to say, this is actually not possible at all. However, what we are actually interested in is not in recovering the noise, but rather in quantifying its impact and then marginalize over it. We therefore start by noting that what enters Eq.(\ref{eq: v0obs})--(\ref{eq: v2obs}) is not the variance of the noise, but rather its ratio with respect to the variance of the convergence. We have therefore to model this quantity, which we do by postulating the following scaling relation
\begin{equation}
{\cal{R}}_k(z_{\rm s}, \theta_{\rm s}) = {\cal{R}}_{k,{\rm ref}}\, \left [ \frac{\sigma_{k,{\rm ref}}}{\sigma_k(z_{\rm s}, \theta_{\rm s})} \right ]\, \left ( \frac{\theta_{\rm s}}{\ang{;2.0;}} 
\right )^{-\alpha_k}
\label{eq: rkmodel}
\end{equation}
that explicitky shows the dependence on the source redshift $z_{\rm s}$ and on the smoothing angle $\theta_{\rm s}$ (in arcmin). This model takes two nuisance parameters for each order $k$, namely the slope $\alpha_k$ of the angle dependence and the value of ${\cal{R}}_k$ at the arbitrary chosen reference point $(z_{\rm s}, \theta_{\rm s}) = (0.3, \ang{;2.0;})$ where the field variance takes the value $\sigma_{k,{\rm ref}}$. Since we expect the noise variance to decrease with the smoothing angle, $\alpha_k$ only takes positive values. The case $\alpha_k = 2$ corresponds to a perfect Gaussian noise.

For the generalized skewness parameters, we follow the same approach and set 
\begin{equation}
S^{(k)}_{\rm N} = \varepsilon_{k}\, S^{(k)}_{\rm ref}\, (\theta_{\rm s}/\ang{;2.0;})^{\beta_k} \, ,
\label{eq: skmodel}
\end{equation}
with $\varepsilon_k$ setting the value of the noise generalized skewness $S^{(k)}_{\rm N}$ to the reference point of the convergence field $S^{(k)}_{\rm ref} = S^{(k)}(z_{\rm s} = 0.3, \theta_{\rm s} = \ang{;2.0;})$, while $\beta_k$ scales the dependence on smoothing radius $\theta_{\rm s}$.  

Summarizing, we end up with two parameters $({\cal{R}}_k, \alpha_k)$ for each variance ratio, plus two parameters $(\varepsilon_k, \beta_k)$ for each generalized skewness, plus the $m_{\kappa}$ multiplicative bias thus summing to a total of ${\cal{N}}_{\rm p} = 13$ nuisance parameters. We stress that Eqs.(\ref{eq: rkmodel}) and (\ref{eq: skmodel}) refer to a fixed $\nu$. If one is interested in using MFs at different S/N thresholds, further nuisance parameters must be added. In this case the total number of nuisance parameters is  ${\cal{N}}_{\rm p} = 12 \times {\cal{N}}_{\nu} + 1$, where ${\cal{N}}_{\nu}$ is the number of threshold values.

\section{Minkowski functionals from simulated maps}

The semi-analytical approach developed in Sect.\,III relates the MFs measured on a realistic convergence map, i.e. one reconstructed from a noisy shear catalogue, with the theoretical MFs computed using leading-order non-Gaussian corrections. The underlying assumption used in Sect.\,III is that the approximate theoretical MFs thus computed differ from the exact theoretical MFs by a calibration factor, i.e., they are linearly related.

We need to test this assumption. In order to do this, we need the ``true" MFs, free from the biases induced by the factor i) to iv) described in Sect.\,III. For this, we will rely on convergence maps with no shape noise or reconstruction bias from the simulated data set that we describe below.

\subsection{The MICE lensing catalogue}

N-body light cone simulations are an ideal tool to build all-sky lensing maps. We rely, in particular, on the the MICE Grand Challenge (MICE-GC) containing about 70 billion dark matter particles in a $(3\, h^{-1} \, {\rm Gpc})^3$ volume \cite{Fosalba2015a,Crocce2015b,Carretero:2017zkw}. The parent simulation is run in a flat $\Lambda$CDM with cosmological parameters set
\begin{displaymath}
(\Omega_{\rm m}, \Omega_{\rm b}, h, n_{\rm s}, \sigma_8) = (0.25, 0.044, 0.70, 0.95, 0.80) \ ,
\end{displaymath}
being $(\Omega_{\rm m}, \Omega_{\rm b}, h, n_{\rm s}, \sigma_8) $ the matter and baryon density parameters, the Hubble constant in units of $100\, \kmsMpc$%$100 \ {\rm km/s/Mpc}$
, the spectral index and the variance of perturbations on the scale $R=8\,\hMpc$ %$R = 8 h^{-1} \ {\rm Mpc}$
, respectively. Galaxies are associated with dark matter haloes using a Halo Occupation Distribution and a Halo Abundance Matching technique \cite{Carretero2015} whose parameters are set to match local observational constraints, such as the local luminosity function \citep{Blanton2003,Blanton2005a}, the galaxy clustering as a function of luminosity and colour \citep{Zehavi2011} and colour-magnitude diagrams \citep{Blanton2005b}.

Given its large volume and fine mass resolution, the MICE-GC simulation also allows an accurate modelling of the lensing observables from upcoming wide and deep galaxy surveys. Following the {\it Onion Universe} approach \cite{Fosalba2008,Fosalba2015b}, all-sky lensing maps are constructed with sub-arcminute scale resolution.  These lensing maps allow to model galaxy lensing properties, such as the convergence, shear, and lensed magnitudes and positions. Tests have been performed to show that the galaxy lensing mocks can be used to accurately model lensing observables down to arcminute scales. We use the MICECAT v2.0 (kindly made available to us by P. Fosalba) which updates the public release MICECAT v1.0 catalog to include less massive and hence lower luminosity galaxies. The galaxy catalog is complete up to $i\sim 24$ for $0 \le z < 1.4$ so that we will hereafter focus our attention on this redshift range only.

\subsection{Measuring Minkowski functionals}

Let $\kappa(\vec{x})$ be a smoothed convergence field with null mean. We define the excursion set $Q_{\nu}$ for a given threshold $\nu$ as the set of points \vec{x} for which the condition $\kappa(\vec{x}) > \nu$ is satisfied. As mentioned in the introduction, the MFs are the area, the perimeter, and the Euler characteristics of $Q_{\nu}$ which, for a pixelized convergence map, can be estimated as \cite{Kratochvil2012} .
\begin{equation}
V_0(\nu) = \frac{1}{A}\int_{A} \,{\Theta(\kappa(\vec{x})-\nu)\; \diff \vec{x}} \; , 
\label{eq: v0nu}
\end{equation}
\begin{equation}
V_1(\nu) = \frac{1}{4 A} \int_{A}{\delta_{\rm D}[\kappa(\vec{x})-\nu]\, \sqrt{\kappa_x^2+\kappa_y^2}\; \diff \vec{x}} \;  , 
\label{eq: v1nu}
\end{equation}
\begin{eqnarray}
V_2(\nu) & = & \frac{1}{2 \pi A} \\
& \times & \int_{A}{{\delta_{\rm D}[\kappa(\vec{x})-\nu]\, \frac{2\kappa_x\kappa_y\kappa_{xy}-\kappa_{x}^2\kappa_{yy}-\kappa_{y}^2\kappa_{xx}}{\kappa_x^2+\kappa_y^2} \; \diff \vec{x}}} \;  , \nonumber
\end{eqnarray}
where $\Theta$ is the Heaviside step function and $\delta_{\rm D}$ is the Dirac function. The integrands are calculated as discrete sums over the pixels. The spatial derivatives of the convergence field are computed via a finite difference scheme with periodic boundary conditions to regularize the borders. The final results are normalized by the area of the maps.

\begin{figure}
	\includegraphics[scale=0.21]{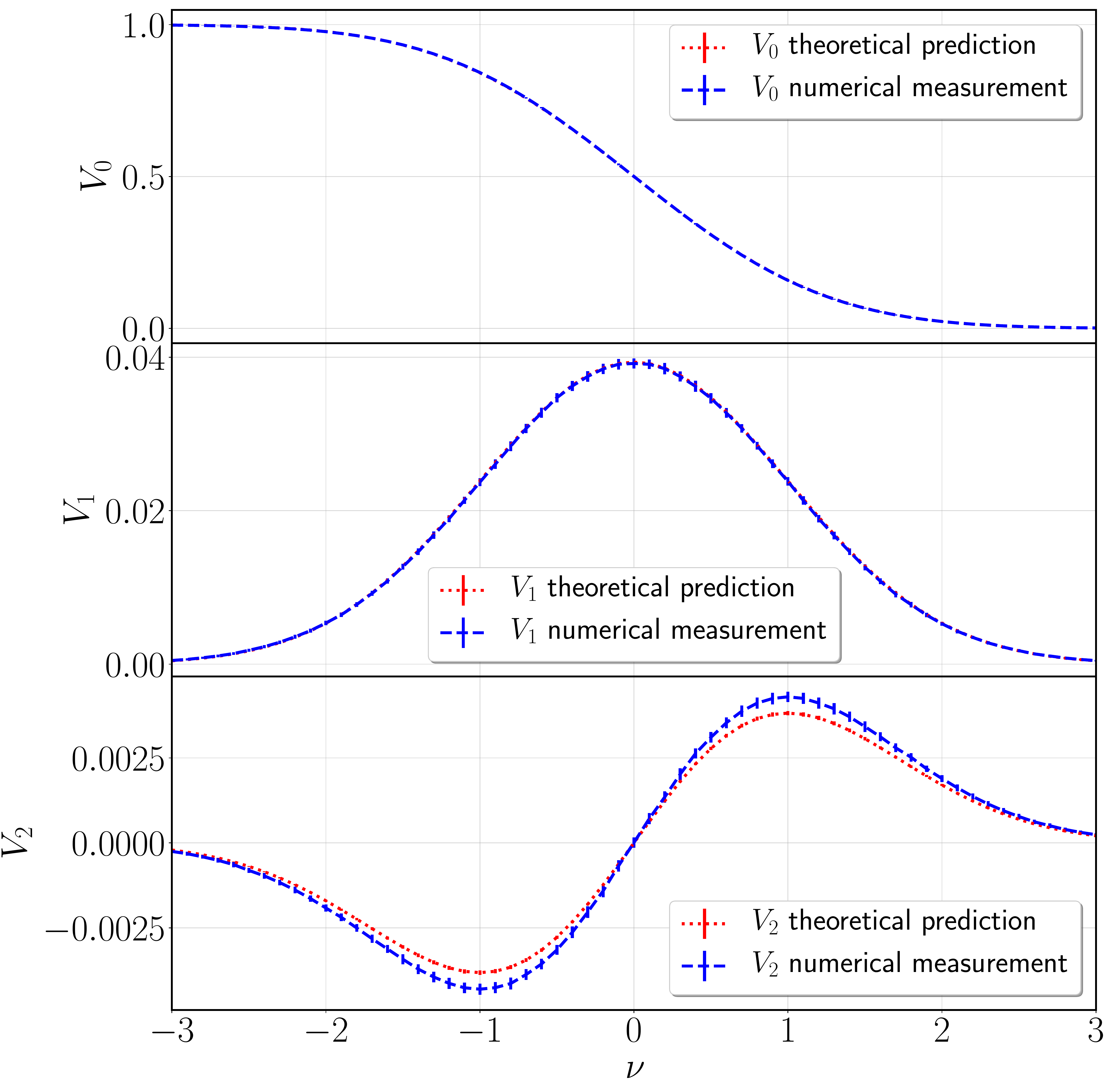}  
	\caption{Minkowski functionals as measured on Gaussian maps contrasted against theoretical predictions.\label{fig: mfgausstest}}
\end{figure}

In order to test whether the code works correctly, we created simulated maps with a Gaussian field and used the above algorithm to measure the MFs. We then compare the measured MFs with the theoretical prediction from Eq.(\ref{eq: GaussMFs}). Fig.\,\ref{fig: mfgausstest} shows that the code correctly recovers the input $(V_0, V_1)$ MFs, but there is a mismatch for $V_2$. This is due to having used the measured $(\sigma_0, \sigma_1)$ to plot the theoretical MFs. We therefore do not consider this deviation as a failure of the code and confidently use it in the following.

\subsection{Convergence maps}
In order to build the convergence maps from shear data, we first cut square patches of $5 \times 5 \, {\rm  deg}^2$ from the MICE catalogue. The patches must be well separated in order to be pseudo-independent realizations of sky.
The final usable area is $\Omega = 3500 \, {\rm  deg}^2$ divided in 140 patches. The area of each patch is large enough to ensure good statistics, but small enough to make deviations from flat sky approximation (used in the theoretical derivation) negligible. The MICECATv2.0 catalogue contains redshift $z$, Right Ascension (RA) and Declination (Dec) each galaxy. We then project these positions onto a tangential plane through a sinusoidal projection. Following \cite{Ludo2013}, we arrange galaxies in approximately square pixels with size \ang{;0.85;}. Such a size ensures a large number of objects in each pixel, (the number density of galaxies is about $n_{\rm g} \simeq 27 \, {{\rm arcmin}^{-2}}$) without compromising the map resolution. We then perturb both shear components by adding random values to mimic the intrinsic ellipticity, thus creating a noisy shear catalogue. We finally reconstruct convergence fields from noisy shear data using the KS method \cite{KS93}.

\subsection{Data covariance matrix}

The ${\cal{N}}_{\rm f}$ convergence maps thus reconstructed are used as input for the estimate of both the fiducial MFs and the associated covariance matrix. To this end, for each given source redshift $z_{\rm s}$, smoothing angle $\theta_{\rm s}$, and threshold $\nu$, we measure the MFs $(V_0, V_1, V_2)$ on the $i$th map. The measured MFs are then given by 
\begin{equation}
V_{n}^{\rm obs}(\nu, z_{\rm s}, \theta_{\rm s}) = \frac{1}{{\cal{N}}_{\rm f}}\, \sum_{i = 1}^{{\cal{N}}_{\rm f}}\,{V_{n}^{(i)}(\nu, z_{\rm s}, \theta_{\rm s})} 
\label{eq: vnobs}
\end{equation}
where $n$ runs from $0$ to $z$. The covariance between two MFs can be written as 
\begin{equation}
{\rm Cov}(V_n, V_m) = \sum_{i = 1}^{{\cal{N}}_{\rm f}}\,
{\frac{[V_{n}^{(i)} - V_{n}^{\rm obs}]\, [V_{m}^{(i)} - V_{m}^{\rm obs}]}{{\cal{N}}_{\rm f} - 1}}
\label{eq: covmap}
\end{equation}
where we stress that $(V_n, V_m)$ can also be evaluated at different $(\nu, z_{\rm s}, \theta_{\rm s})$. This is the error for the MFs evaluated on a single $25 \,{ \rm deg}^2$ map. The observed covariance is the error on the mean. It can be obtained from Eq.(\ref{eq: covmap}) by scaling down by a factor $\Omega/25$ to account for the number of patches in different survey areas $\Omega$. 

It is worth noting that such a scheme implicitly assume that a single $25 \,{ \rm deg}^2$ subfield is fully representative of the total survey area so that one can apply a sort of ergodicity theorem considering different patches as multiple realizations of the same universe. This is a common assumption and we do not question its validity here. We nevertheless note that this issue needs to be reconsidered should masking be applied. Since we do not include masking in the present analysis, we can take as granted the validity of this popular approach.

\section{Validating the matching}

Eqs.(\ref{eq: v0obs})--(\ref{eq: v2obs}) provide a way to compute the expected MFs values from cosmological- and noise-related quantities, by modelling the various biases with a simple multiplicative factor $m_k$. In turn, the amplitude and smoothing angle dependence of these noise-related quantities (variance ratios and generalized skewness quantities) are also modelled in Eq.(\ref{eq: rkmodel})--(\ref{eq: skmodel}), introducing more nuisance parameters, but avoiding the need of modelling the properties of the noise field. The assumption of a multiplicative bias, although realistic, must be validated against simulated data. If such linear scaling is indeed found, then we will be able to rely on the linear-order approximation to estimate the theoretical MFs for a given set of cosmological parameters, without introducing any systematic error. Moreover, the match with the simulated data allows us to infer the values of the nuisance parameters and to investigate which values of the S/N threshold $\nu$ give the more reliable MFs.

To this end, we implement a two step procedure that we will describe in detail below. As input, we preliminary estimate MFs from the simulated maps for four different source redshifts\footnote{To this end, we perform the map reconstruction selecting from the full catalogue only galaxies with redshift $z$ in the range $(z_{\rm s} - 0.05, z_{\rm s} + 0.05)$. This turns out to be a good compromise between the need for a small range, avoiding non constant MFs, and a large number of sources to guarantee decent statistics.}, namely $z_{\rm s} = (0.3, 0.6, 0.9, 1.2)$, and four different values of the smoothing angle $\theta_{\rm s} = (2.0, 6.0, 10.0, 14.0) \, {\rm arcmin}$. We consider both $\nu = 2$ and $\nu = 3$ as S/N thresholds. Note that $\nu = 1$ is not used since ${\cal{H}}_n(\nu = 1) = 0$ for all $n$, so $\delta V_{n}^{(2)} = 0$ and the theoretical MFs reduce to the Gaussian ones. This is a consequence of our choice to halt the perturbative MF expansion at first order in $\sigma_0$. In this case, non-Gaussianities only appear at second order in $\sigma_0$ which require modelling the trispectrum. Given the considerable theoretical uncertainties, we choose to avoid this task cannot consider the $\nu = 1$ case.

\subsection{Calibrating the relation for $V_0$}

Although, for a given $\nu$, the total number of nuisance parameters in our calibration procedure is ${\cal{N}}_{\rm p} = 13$, not all of them enters all the relations. Indeed, Eq.(\ref{eq: v0obs}) shows that, in order to predict the first MF, one only needs 5 out of the 13 parameters, namely the multiplicative bias $m_{\kappa}$, the $({\cal{R}}_0, \alpha_0)$ parameters that set the scaling of the ratio between noise and convergence variances, and $(\varepsilon_0, \beta_0)$ that fix the behaviour of the noise generalized skewness at the lowest order. With ${\cal{N}}_{\rm p} = 5$ parameters and ${\cal{N}}_{\rm d} = 4 \times 4 = 16$ computed $V_{0,{\rm obs}}$ values for all the $(z_{\rm s}, \theta_{\rm s})$ combinations, one could adopt a brute force approach minimizing a $\chi^2$-like  merit function over a 5D grid. However, degeneracy among the parameters and the non-negligible statistical errors on the measured MFs make quite severe the risk of getting stuck in local minima.  

To reduce this possibility, we have therefore developed an efficient procedure to determine the nuisance parameters entering the $V_0$ calibration. To this end, let us first note that Eq.(\ref{eq: v0obs}) may be easily solved with respect to $S_{\rm N}^{(0)}$ as
\begin{equation}
S_{\rm N}^{(0)} = \frac{6\, [(1 + m_{\kappa}^2) + {\cal{R}}_{0}^{2}]^{3/2}}{\sigma_0^2\, {\cal{H}}_{2}(\nu)} \,
\frac{\Delta V_{0}^{\rm obs}}{{\cal{R}}_{0}^4} - \frac{(1 + m_{\kappa})^3\, S^{(0)}}{{\cal{R}}_0^4}
\label{eq: s0nsol}
\end{equation}
with
\begin{equation}
\Delta V_{0}^{\rm obs} = \frac{\sqrt{2 \pi}\, V_{0,{\rm obs}}}{\exp{(-\nu^2/2)}} - {\cal{H}}_{-1}(\nu) \; . 
\label{eq: deltav0obs}
\end{equation}
Starting from this consideration, we then proceed as follows for a fixed $\nu$ value. 

\begin{enumerate}
\item{For each of 16  $(z_{\rm s}, \theta_{\rm s})$ cases we compute 16 $V_{0}^{\rm obs}$ functionals and we use Eq.(\ref{eq: s0nsol}) to get $S^{(0)}_{N}$, fixing the remaining nuisance parameters $(m_{\kappa}, {\cal{R}}_{0,{\rm ref}}, \alpha_0)$. By propagating the errors on $V_{0,{\rm obs}}$, we also get the uncertainty on $S^{(0)}_{\rm N}$.} 

\item{For each $\theta_{\rm s}$ we should expect identical $S^{(0)}_{\rm N}$ for the 4 $z_{\rm s}$ cases as postulated in Eq.(\ref{eq: skmodel}). However, this is not the case because the dependence on the redshift enters in Eq.(\ref{eq: s0nsol}) through ${\cal{R}}_{0}$. We consider then a $\hat{S}^{(0)}_N(\theta_{\rm s})$ which is estimated as the error weighted mean of the values from the four redshift cases. We prepare a table with the $\theta_{\rm s}$ and $\hat{S}^{(0)}_N(\theta_{\rm s})$ values.}

\item{We fit Eq.(\ref{eq: skmodel}) to the table, thus getting an estimate of the parameters $(\varepsilon_0, \beta_0)$ for the given $(m_{\kappa}, {\cal{R}}_{0,{\rm ref}}, \alpha_0)$. Note that this will not be the best fit to the actual $V_{0,{\rm obs}}$ data since we have minimized the fit to the average $S^{(0)}_{N}$ values. }

\item{We now repeat steps 1--3 over a fine grid in the 3D space $(m_{\kappa}, {\cal{R}}_{0,{\rm ref}}, \alpha_0)$ assigning to each set of parameters a merit function defined as 
\begin{displaymath}
\chi^2 = \sum_{i}\,{\left [ \frac{V_{0}^{\rm meas}(z_{{\rm s},i}, \theta_{{\rm s},i}) - V_{0,{\rm obs}}(z_{{\rm s},i}, \theta_{{\rm s},i}, \vec{p_0})}
{\sigma_{i}^{\rm meas}} \right ]^2}
\end{displaymath}
where $V_{0}^{\rm meas}(z_{{\rm s},i}, \theta_{{\rm s},i})$ is the measured value for the $i$th $(z_{\rm s}, \theta_{\rm s})$ combination, $\sigma_{i}^{\rm meas}$ the statistical uncertainty, $\vec{p_0} = (m_{\kappa}, {\cal{R}}_{0,{\rm ref}}, \alpha_0, \varepsilon_{0}, \beta_0)$ are the nuisance parameters with the last two ones fixed as function of the others as in step 3, and the sum runs over all the $(z_{\rm s}, \theta_{\rm s})$ data set.}

\end{enumerate}
Although not completely statistically motivated (since we are not performing a full $\chi^2$ minimization\footnote{For this same reason, we do not give the number of degrees of freedom. This should be computed as ${\cal{N}}_{dof} = {\cal{N}}_d - {\cal{N}}_p$ with ${\cal{N}}_d$ the number of data used and ${\cal{N}}_p$ that of calibration parameters. However, we only fit two of them over a grid defined by the other three parameters so that we actually perform a two steps minimization with a varying numbers of degrees of freedom.}), this scheme nevertheless allows us to find the best set of nuisance parameters that minimize the scatter between the measured MFs and the predicted ones. Indeed, for $\nu = 2$, we find that setting
\begin{displaymath}
m_{\kappa} = 0.144 \ \ , {\cal{R}}_{0,{\rm ref}} = 12.59 \ \ , \ \  \alpha_0 = 1.15  \ \ ,
\end{displaymath}
\begin{displaymath}
\varepsilon_0 =  505.129 \ \ , \ \ \beta_0 = 1.29 
\end{displaymath}
gives $\chi^2/N_{\rm dof} = 0.029$ indicating a good match between theoretical predictions and measurements. One could wonder whether such a small reduced $\chi^2$ is actually an evidence of overestimated errors instead of the model fitting the (simulated) data. We have therefore also computed the percentage residuals $\Delta_0 = 1 - V_{0,{\rm obs}}/V_{0,{\rm meas}}$ with the subscript obs (meas) labelling the predicted (measured) values of $V_0$. We find a median value $med(\Delta_0) = 0.8\%$ and a root mean square $rms(\Delta_0) = 7.8\%$ which make us confident that the model indeed matches the simulated measurements.

This can also be seen as an evidence that our fitting procedure indeed works. On the contrary, for $\nu = 3$, we now get $med(\Delta_0) = -16\%$ and $rms(\Delta_0) = 26\%$ which are clearly unsatisfactory. We argue that the problem actually goes back to Eq.(\ref{eq: mfsum}) and our decision to halt the expansion at the first order in $\sigma_0$. This is a resonable approximation as far as $\delta V_3(\nu) \ll \delta V_2(\nu)$ which is likely not the case for $\nu = 3$. 

\subsection{Calibrating the relations for $(V_1, V_2)$}

The same kind of procedure can be implemented to validate Eqs.(\ref{eq: v1obs})-(\ref{eq: v2obs}) for the observable $(V_1, V_2)$ MFs. The key difference, however, is that the coupling between different orders induced by the noise makes all the 13 nuisance parameters enter these relations. As a consequence, it is no longer possible to solve for, e.g., $S^{(1)}_{\rm N}$ from $V_{1,obs}$ only, but rather one has to consider $(V_{1,{\rm obs}}, V_{2,{\rm obs}})$ to get both $S^{(1)}_{N}$ and $S^{(2)}_{N}$. We therefore implement a similar procedure as the one already described for the zero\,-\,order MF, but with some differences that we highlight below.

\begin{enumerate}
\item[i.]{We fix the nuisance parameters $(m_{\kappa}, {\cal{R}}_{0,{\rm ref}}, \alpha_0, \varepsilon_0, \beta_0)$ to the best fit values found when calibrating the relation for $V_0$. This reduces the dimensionality of the problem.}

\item[ii.]{For given values of $({\cal{R}}_{1,{\rm ref}}, \alpha_1, {\cal{R}}_{2,{\rm ref}}, \alpha_2)$, we solve Eqs.(\ref{eq: v1obs})-(\ref{eq: v2obs}) with respect to $S^{(1)}_{\rm N}$ and $S^{(2)}_{\rm N}$. Note that we propagate the uncertainties solving the system of equations after sorting the $(V_1, V_2)$ from a Gaussian distribution centred on the measured value and with width set by the corresponding errors.}

\item[iii.]{We repeat steps 2 and 3 described before, but now for the 1st and 2nd order generalized noise skweness parameters which give us estimates of $(\varepsilon_1, \beta_1, \varepsilon_2, \beta_2)$.}

\item[iv.]{We find the best fit $({\cal{R}}_{1,{\rm ref}}, \alpha_1, {\cal{R}}_{2,{\rm ref}}, \alpha_2)$ parameters by minimizing a $\chi^2$ function obtained by summing up in quadrature the residuals for $V_1$ and $V_2$.}

\end{enumerate}
For $\nu = 2$, we find as best fit parameters
\begin{displaymath}
{\cal{R}}_{1,{\rm ref}} = 2.52 \ \ , \ \ \alpha_1 = 2.0 \ \ , 
\end{displaymath}
\begin{displaymath}
\varepsilon_1 = -10109.9 \ \ , \ \ \beta_1 = 1.70 \ \ ,
\end{displaymath}
\begin{displaymath}
{\cal{R}}_{2,{\rm ref}} = 2.52 \ \ , \ \ \alpha_2 = 3.6 \ \ , 
\end{displaymath}
\begin{displaymath}
\varepsilon_2 = -283205 \ \ , \ \ \beta_2  = 1.86 \ \  .
\end{displaymath}
Note that the surprisingly large values of $\varepsilon_1$ and $\varepsilon_2$ are partly a consequence of having scaled the noise generalized skewness moments with respect to the convergence ones. Since $S^{(1)}_{\rm N}$ and $S^{(2)}_{\rm N}$ are related to the spatial derivatives of the noise, it is expected that they are quite larger than the convergence ones because of the quick random variations of the noise over the field area. We therefore expect $S^{(1)}_{\rm N} \gg S^{(1)}$ and $S^{(2)}_{\rm N} \gg S^{(2)}$ which is what the large values we find are confirming.

We again find that the fit performs quite well with $\chi^2/N_{\rm dof} = 0.56$, while the percentage residuals read
\begin{displaymath}
med(\Delta_1) = 14.1\% \ \ , \ \ RMS(\Delta_1) = 16.5\% \ \ ,
\end{displaymath}
\begin{displaymath}
med(\Delta_2) = 7.1\% \ \ , \ \ RMS(\Delta_2) = 20.1\% \ \ ,
\end{displaymath}
%S
which are again evidence that the calibration relations work quit well in matching theoretical and measured MFs. 

For $\nu = 3$, the calibration procedure dramatically fails with median and root mean square of the residuals unacceptably large. We will therefore not consider anymore the case $\nu = 3$ in the following fixing henceforth $\nu = 2$ as the threshold to measure MFs on the convergence map. 

\section{Fisher matrix forecasts}

The calibration procedure described and validated above have convincingly shown that it is possible to match theoretical and measured MFs even if the MFs series expansion is halted at the lowest order. The cost to pay is the addition of nuisance parameters which correct for the overall impact of missing higher-order terms, imperfect reconstruction from the shear field, and noise in the ellipticity data. These quantities could be calibrated against simulations mimicking as close as possible the details of the data set to be used. Actually, such a procedure can only give constraints on the nuisance parameters rather than fixing their values with infinite precision. It is therefore a safer strategy to treat them as unknowns, adding them to the list of parameters to determine from the fit to the data. The full parameters vector \vec{p} will be the union of the cosmological one, with seven components\footnote{Fiducial values of the cosmological parameters are the same as the input ones for the MICE simulation.}
\begin{displaymath}
\vec{p}_{\rm c} = \{\Omega_{\rm m}, \Omega_{\rm b}, w_0, w_a, h, n_{\rm s}, \sigma_8 \} \ ,
\end{displaymath}
and the nuisance one, $\vec{p}_{\rm n}$, whose components depend on which combination of MFs one decides to use. On the one hand, one could naively argue that by using more MFs, more information is included and the constraints would be stronger. However, using more MFs, introduces a higher number of nuisance parameters and hence more parameters to constraints. We will therefore investigate different options looking for the one that gives the strongest constraints on the cosmological parameters.

Given the large number of parameters (from 12 if $V_0$ only is used to 20 if all MFs are used), we do not expect MFs alone to be able to put severe constraints on all of them. We will therefore consider MFs as a way to improve the constraints coming from the standard second-order statistics. The total Fisher matrix will then be the sum of the MFs one and the shear tomography contribution computed as described in Appendix B of \cite{Vicinanza2018HOM}. Note that we are here assuming that MFs and the shear power spectra are independent probes. Although they both come from the same observable (the ellipticity of galaxies), we expect that the many steps needed to go from raw data to MFs cancel any correlation with the shear data thus justifying our approach as a reasonable first order approximation. 

\subsection{Fisher matrix formalism}

We rely on the Fisher matrix formalism \cite{Tegmark1997} and compute the Fisher matrix elements given by
\begin{equation}
F_{ij} = \left \langle - \frac{\partial^2 {\ln{\cal{L}}}(\vec{p})}{\partial p_i \partial p_j} \right \rangle \; .
\label{eq: fijdef}
\end{equation} 
The averaging is approximated by considering the likelihood value at the fiducial model and $\vec{p}_{\rm fid}$. Assuming a multivariate Gaussian, the likelihood function reads
\begin{equation}
-2 \ln{{\cal{L}}(\vec{p})} \propto
(\vec{D}_{\rm obs} - \vec{D}_{\rm th})^T\, {\bf Cov}^{-1}\, (\vec{D}_{\rm obs} - \vec{D}_{\rm th}) \; ,
\label{eq: deflike}
\end{equation}
where $\vec{D}_{\rm obs}$ is the data set vector, and $\vec{D}_{\rm th}$ is its theoretical counterpart evaluated as described before. The Fisher matrix elements therefore read
\begin{equation}
F_{ij} = \frac{\partial \vec{D}_{\rm th}}{\partial p_i}\, {\bf Cov}^{-1}\, \frac{\partial \vec{D}_{\rm th}}{\partial p_j} \; .
\label{eq: fijhom}
\end{equation}
The data covariance matrix plays a key role. We split it in two terms 
\begin{equation}
{\bf Cov} = {\bf Cov}^{\rm obs} + {\bf Cov}^{\rm sys}
\label{eq: defcovtot}
\end{equation}
with the first term gives the statistical errors, while the second one is introduced to take into account the inaccuracies in the calibration procedures, i.e., the fact that we can predict MFs from theory only up to a scatter quantified by the RMS of percentage residuals. This covariance matrix due to the systematic effects is diagonal so that its elements read
\begin{equation}
{\rm Cov}_{ij}^{\rm sys} = \rho_{\rm RMS}(i)\, \rho_{\rm RMS}(j)\, D_{\rm obs}(i)\, D_{\rm obs}(j) \delta^{\rm K}_{ij}
\label{eq: covijsys}
\end{equation}
with $\delta^{\rm K}_{ij}$ the Kronecker $\delta$, and $\rho_{\rm RMS}(i)$ the RMS of the percentage residuals for the MF corresponding to the $i$th element  of the data vector $\vec{D}_{\rm obs}$. What actually enters the likelihood is the precision matrix, i.e., the inverse of the covariance matrix. However, it is known that this quantity may be biased if estimated from a not large enough number of independent realizations. Following \cite{Hartlap2007}, we compute it as 

\begin{figure*}
\centering
\includegraphics[width=5.5cm]{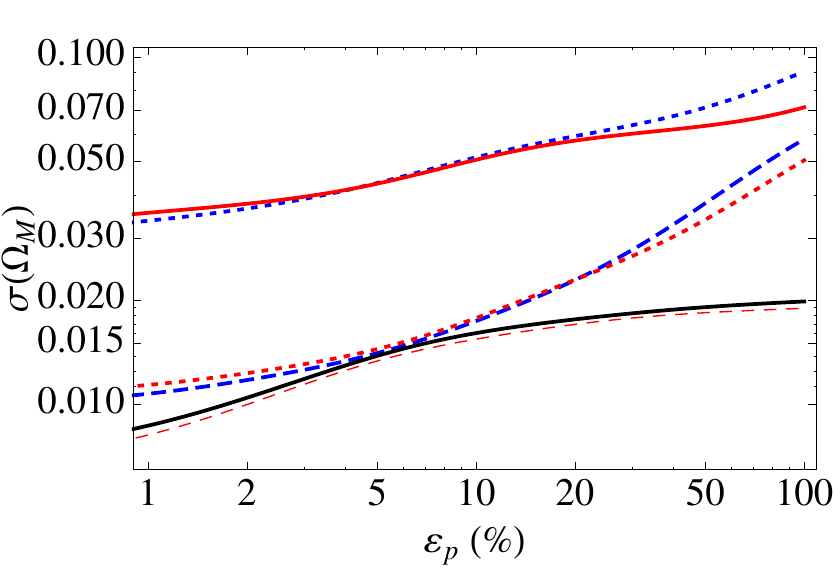}
\includegraphics[width=5.5cm]{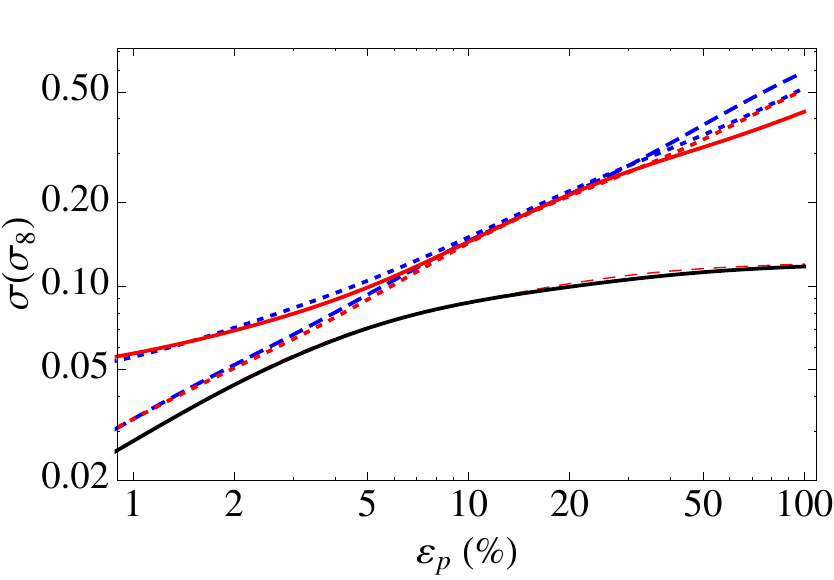}
\includegraphics[width=5.5cm]{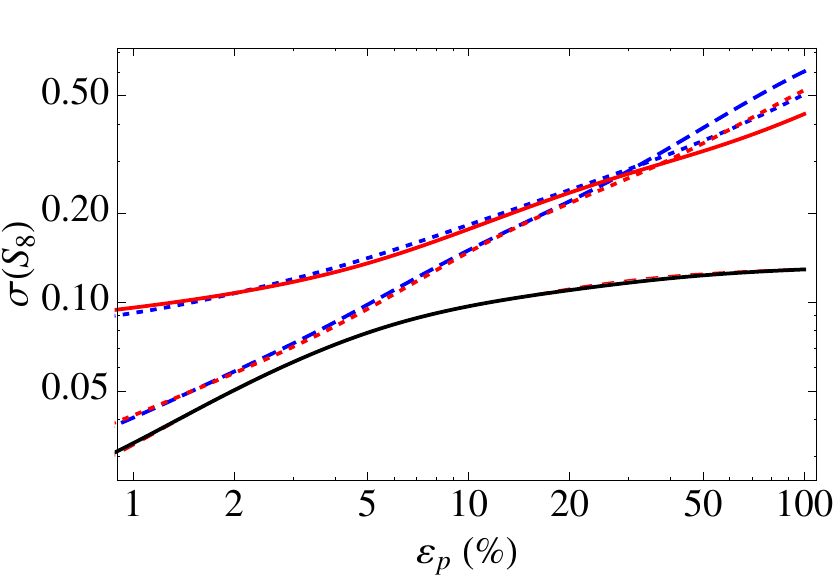} 
\caption{Constraints on $(\Omega_{\rm m}, \sigma_8, S_8)$ from MFs only as a function of the prior $\varepsilon_P$ on the nuisance parameters for a survey with area $\Omega = 3500\, {\rm deg}^2 $. Hereafter, blue dotted, blue dashed, red solid, red dotted, red dashed, black solid lines refer to the case when $\vec{D}_1$, $\vec{D}_2$, $\vec{D}_0 \cup \vec{D}_1$, $\vec{D}_0 \cup \vec{D}_2$, $\vec{D}_1 \cup \vec{D}_2$, $\vec{D}_0 \cup \vec{D}_1 \cup \vec{D}_2$ data are used.\label{fig: oms8mfonly}}
\end{figure*}
\begin{equation}
{\bf Cov}^{-1} = \frac{{\cal{N}}_{\rm f} - {\cal{N}}_{\rm d} - 2}{{\cal{N}}_{\rm f} - 1}\, ({\bf Cov}^{\rm obs} + {\bf Cov}^{\rm sys})^{-1}
\label{eq: covinv}
\end{equation}
where the multiplicative term corrects for the finite number of realizations ${\cal{N}}_{\rm f} = \Omega/25$ used to estimate the covariance of the ${\cal{N}}_{\rm d}$ dimensional data vector. The full data vector is constructed as follows
\begin{displaymath}
\vec{D}_{\rm obs} = \vec{D}_{0} \cup \vec{D}_1 \cup \vec{D}_2
\end{displaymath}
with
\begin{eqnarray}
\vec{D}_n & = & \left \{V_n(0.3, \ang{;2;}), V_{n}(0.3, \ang{;6;}), V_{n}(0.3, \ang{;10;}), V_{n}(0.3, \ang{;14;}) \right \} \nonumber \\
 & \cup & \left \{V_n(0.6, \ang{;2;}), V_{n}(0.6, \ang{;6;}), V_{n}(0.6, \ang{;10;}), V_{n}(0.6, \ang{;14;}) \right \} \nonumber \\
 & \cup & \left \{V_n(0.9, \ang{;2;}), V_{n}(0.9, \ang{;6;}), V_{n}(0.9, \ang{;10;}), V_{n}(0.9, \ang{;14;}) \right \} \nonumber \\
 & \cup & \left \{V_n(1.2, \ang{;2;}), V_{n}(1.2, \ang{;6;}), V_{n}(1.2, \ang{;10;}), V_{n}(1.2, \ang{;14;}) \right \} \nonumber \; . 
\end{eqnarray}
Here $V_n(z, \theta_{\rm s})$ is the MF of order $n$ estimated from the map made from sources at redshift $z$ after smoothing with a Gaussian filter of width $\theta_{\rm s} \ {\rm arcmin}$ and considering only the regions with $\nu > 2$. We will consider 6 different cases including only one, two or the three $\vec{D}_n$ terms in the data vector.

As mentioned above, the nuisance parameters can be constrained by fitting the theoretical model to simulated data mimicking as close as possible the actual maps. We can add this information to the fit rewriting the Fisher matrix elements as 
\begin{equation}
F_{ij} = \frac{\partial \vec{D}_{\rm th}}{\partial p_i}\, {\bf Cov}^{-1}\, \frac{\partial \vec{D}_{\rm th}}{\partial p_j} + \pi_{ij} \ ,  
\label{eq: fijhomprior}
\end{equation}
where $\pi_{ij}$ are the elements of the prior matrix. This is a diagonal matrix whose first seven elements are set to zero (since we do not want to put any prior on the cosmological parameters), while the remaining elements are set to $1/\sigma_i^2$ with $\sigma_i$ quantifying our belief in the prior knowledge of the nuisance parameters. Setting, e.g., $\sigma_i = 0.1 p_{\rm fid}$ means we are assuming that the true value of the $i$th component of the nuisance parameters vector lies within $10\%$ of the corresponding fiducial one. In order to reduce the quantities to change in the analysis, we will set $\sigma_i = \varepsilon_p \times p_{i}^{fid}$ and explore how constraints change as a function of $\varepsilon_p$. 
\begin{figure*}
\centering
\includegraphics[width=5.5cm]{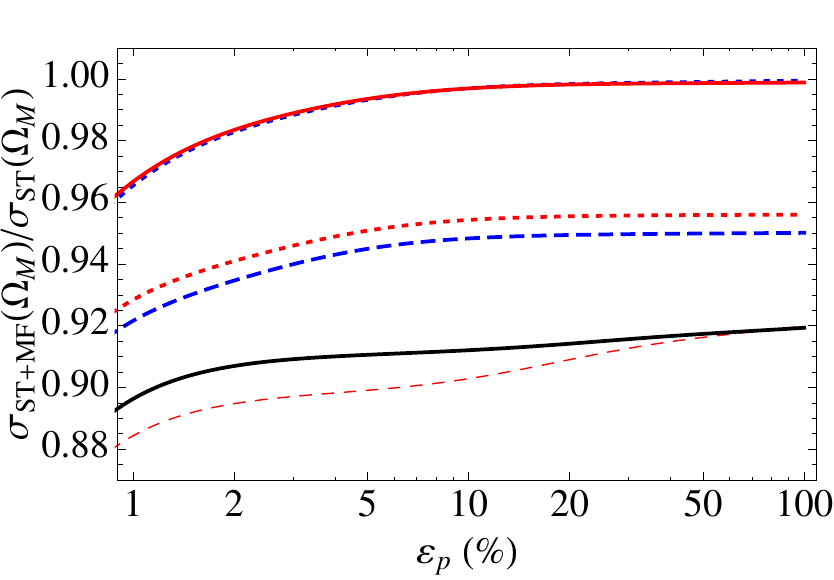}
\includegraphics[width=5.5cm]{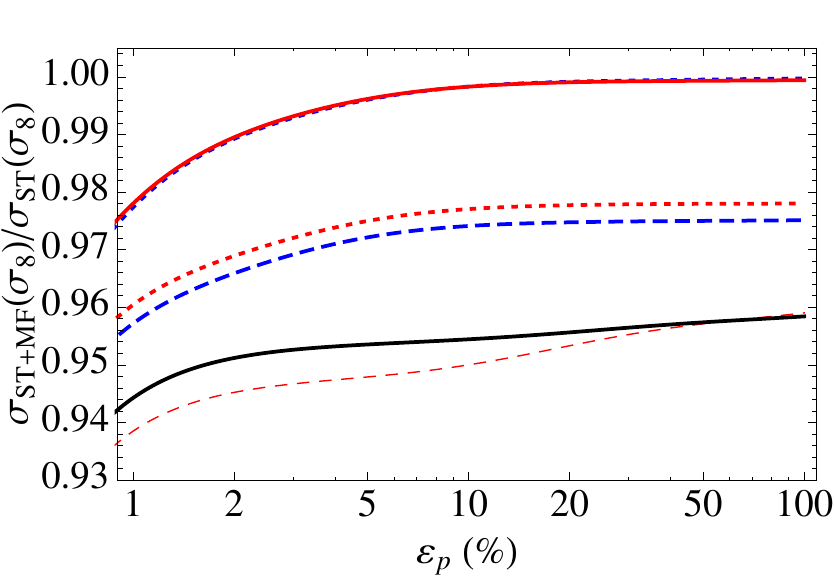}
\includegraphics[width=5.5cm]{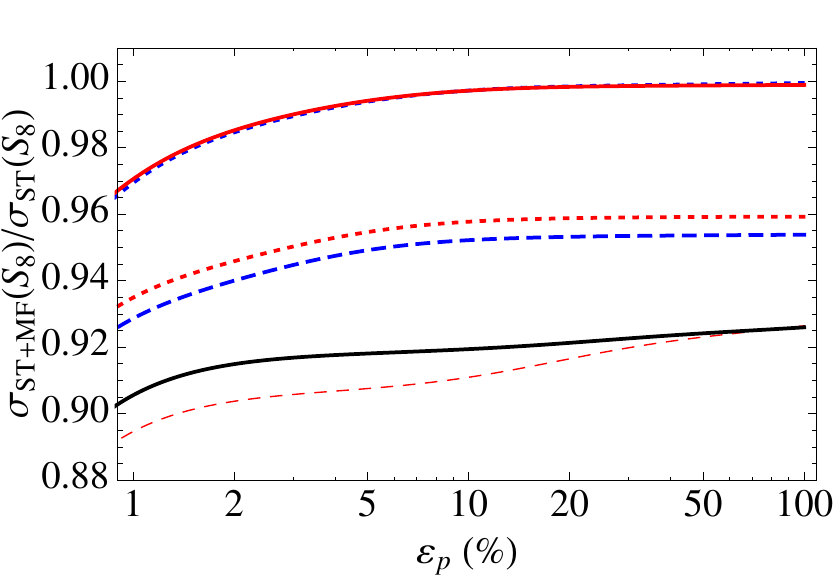} 
\caption{Constraints on $(\Omega_{\rm m}, \sigma_8, S_8)$ from the joint use of MFs and shear tomography. Color code as in Fig.\,\ref{fig: oms8mfonly}. \label{fig: oms8mf6B}}

\end{figure*}

\subsection{Constraints on $(\Omega_{\rm m}, \sigma_8)$ only}

As a preliminary case, we do not let all cosmological parameters free, but we follow the common practice of fixing all of them to the fiducial values leaving only $(\Omega_{\rm m}, \sigma_8)$ and the nuisance parameters free to vary. For completeness, we first look at the constraints from MFs only. After marginalizing over $\vec{p}_{\rm n}$, we get the errors shown in Fig.\,\ref{fig: oms8mfonly} where we have set $\Omega = 3500 \, {\rm deg}^2$ since this is the total simulated area we have used to calibrate the MFs. We also consider the quantity $S_8 = (\Omega_{\rm m}/0.3)^{1/2}\, \sigma_8$ which approximately follows the $\Omega_{\rm m}$--$\sigma_8$ degeneracy line.

It is instructive to see how the six lines separate in three distinct groups of two lines. The two lines in the pairs only differ by the inclusion of the $\vec{D}_0$ data set. This means, for instance, one gets essentially the same results using $\vec{D}_1$ only (blue dotted line) or $\vec{D}_0 \cup \vec{D}_1$ (red solid). We can therefore conclude that the zeroth-order MF does not carry any additional information that is not already included in the MFs of order 1 and 2. Such result could have been anticipated looking at Eqs.(\ref{eq: v0obs})--(\ref{eq: v2obs}) which show that all the quantities entering $V_0(z, \theta_{\rm s})$ also enter the higher-order MFs so that the $\vec{D}_0$ data set do not add further information. This is not the case when $\vec{D}_1$ and $\vec{D}_2$ are combined (red dashed and black solid lines) since $V_1$ and $V_2$ are functions of different moments and generalized skewness of the convergence distribution.

It is worth wondering whether MFs indeed improve the constraints with respect to the case where shear tomography only is used. To this end, we plot in Fig.\,\ref{fig: oms8mf6B} the ratio between the forecasted error from joint MFs and shear and the one from shear only. Smaller ratios mean that the addition of MFs is more useful. As seen in Fig.(\ref{fig: oms8mf6B}), the inclusion of MFs data does not bring a large improvement, with the errors being reduced by no more than a factor of $10\%$ in the more favourable case. Although this result can be discouraging, one should keep in mind that we have fixed most of the cosmological parameters thus eliminating the most severe degeneracies in shear tomography. As a consequence, there is no need for any help in breaking degeneracies and hence MFs seem can not improve the constraints significantly. 

It is nevertheless worth noting that the MFs data set best suited to improve the constraints is not the one that include all MFs, but the $\vec{D}_1 \cup \vec{D}_{2}$ combination. As a general remark, the inclusion of the $\vec{D}_0$ data set worsens the constraints, pointing at a problem in using the zeroth order MF. As already mentioned before, $V_0$ only depends on the variance $\sigma_0$ and $S^{(0)}$, which are both related to second-order statistics as the shear tomography itself. As such, there is no additional information in $V_0$. On the contrary, including $V_0$ adds a contribution to the systematics covariance matrix thus lowering the overall MFs S/N ratio. 

\begin{figure*}
\centering
\includegraphics[width=7.5cm]{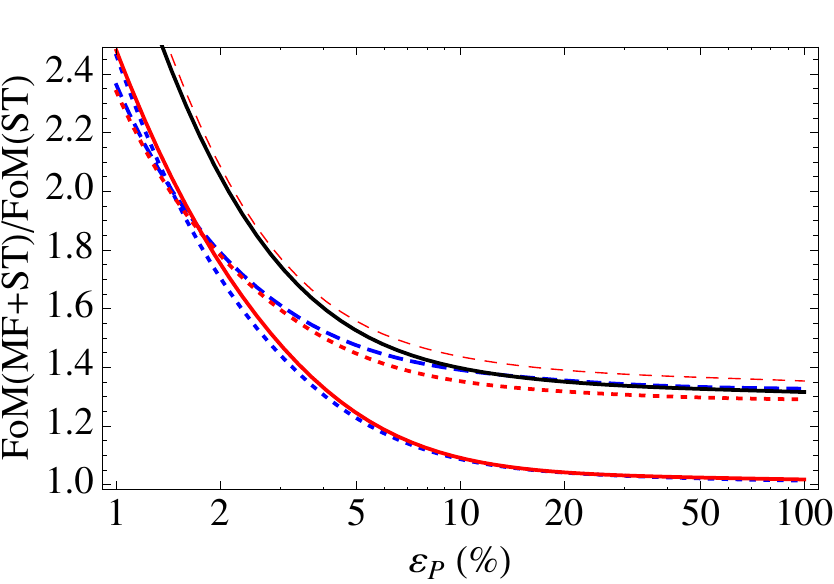}
\includegraphics[width=7.5cm]{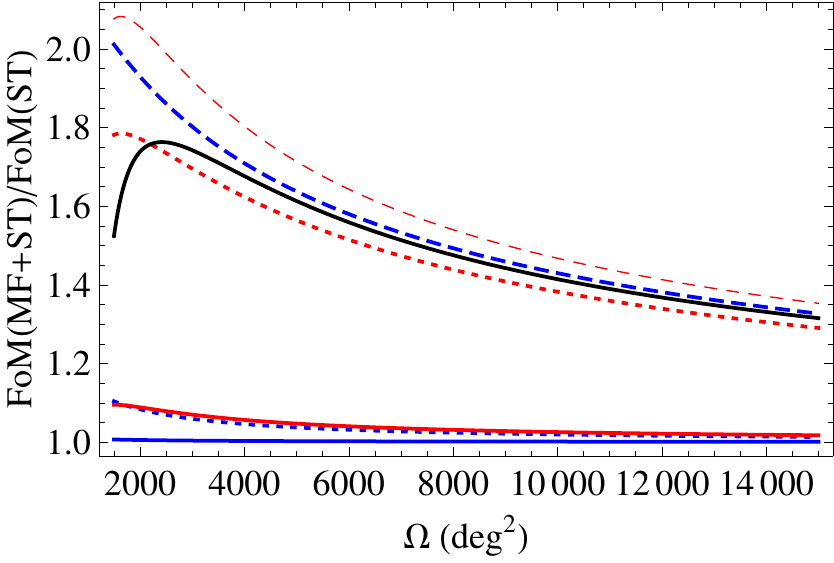}
\caption{FoM ratio as a function of the prior $\varepsilon_P$ on the nuisance parameters and the total survey area $\Omega$. In the left panel, we fix $\Omega = 15\,000 \, {\rm deg}^2$, while it is $\varepsilon_P = 100\%$ in the right panel. Color code is the same as in Fig.\,\ref{fig: oms8mfonly}, but we add, for completeness, a green line (in the right panel only) for the case when the MFs $\vec{D}_0$ data set is used. \label{fig: fomboost}}
\end{figure*}

\subsection{MFs and Figure of Merit}

We now come to the core result of this paper presenting the improvement in the lensing dark energy FoM by the joint use of both shear tomography (hereafter ST) and MFs. Fig.\,\ref{fig: fomboost} shows the ratio between the dark energy FoM with and without MFs included for different data sets. 

Let us first discuss the dependence on the prior $\varepsilon_P$ on the nuisance parameters looking at the left panel where the survey area is fixed to $\Omega = 15\,000 \, {\rm deg}^2$ as for a Euclid-like survey\footnote{Note that this is not the same as the Euclid survey since the redshift distribution of the MICE data set is much different.}. As expected, the smaller is the uncertainty $\varepsilon_P$, the larger is the increase int the FoM, with an improvement larger than a factor 2 (even for the less favourable cases) when $\varepsilon_P < 1\%$. Needless to say, such an accuracy in the prior knowledge of the MFs nuisance parameters is hard to achieve if not unrealistic at all. However, the FoM ratio remains constant for $\varepsilon_P > 10\%$, suggesting that, in order for the MFs to be useful probes, the nuisance parameters do not need to be known with very high precision. Mathematically, this is a consequence of the prior matrix elements $\pi_{ij}$ becoming subdominant with respect to the MFs and ST ones in the total Fisher matrix. Physically, this can be explained noting that the improvement in the FoM is due to MFs partially breaking some of the ST degeneracies. As soon as the MFs nuisance parameters are known with sufficient precision to trigger this mechanism, it becomes irrelevant the exact values of the calibration quantities. 

It is remarkable that, even with $\varepsilon_P = 100\%$ and a survey area $\Omega = 15\,000 \, {\rm deg}^2$, the inclusion of MFs of 2nd order (i.e., $V_2$ data) make the FoM increase by $25 - 30\%$ regardless of $V_0$ and $V_1$ being used or not. This can be qualitatively explained noting that $V_2$ depends on both the derivative and the Laplacian of the convergence field hence probing up to the fourth derivative of the lensing potential. As a consequence, $V_2$ is extremely sensitive to the details of the growth of structure and hence to the cosmological parameters. It is, however, worth noticing that $V_1$ also plays a non-neglible role as it can be seen by comparing the blue and red lines of the $\vec{D}_2$ only and $\vec{D}_1 \cup \vec{D}_2$ cases. The blue line is always below the red one although the difference becomes less important as the survey area increases. This is also seen in the right panel of Fig.\,\ref{fig: fomboost}, showing the ratio $FoM(MF + ST)/FoM(ST)$ as a function of the survey area $\Omega$ in the conservative case $\varepsilon_P = 100\%$. Indeed, the red line ($\vec{D}_1 \cup \vec{D}_2$) stays always on top of the blue one ($\vec{D}_2$ only) thus highlighting again the importance of including $V_1$ and $V_2$ data although both MFs depend on the same quantities. We therefore recommend using both data sets in order to maximize the impact of MFs on the total FoM.

On the contrary, we find again that adding $V_0$ decreases the FoM instead of increasing it, as it can be seen by noting that the black line stays below the blue and red ones. As clearly shown from the green line, $V_0$ data do not add any information which is not already coded into the ST so that its presence only adds noise to the MFs covariance matrix thus decreasing, the overall FoM. 

\begin{figure*}
\centering
\includegraphics[width=5.5cm]{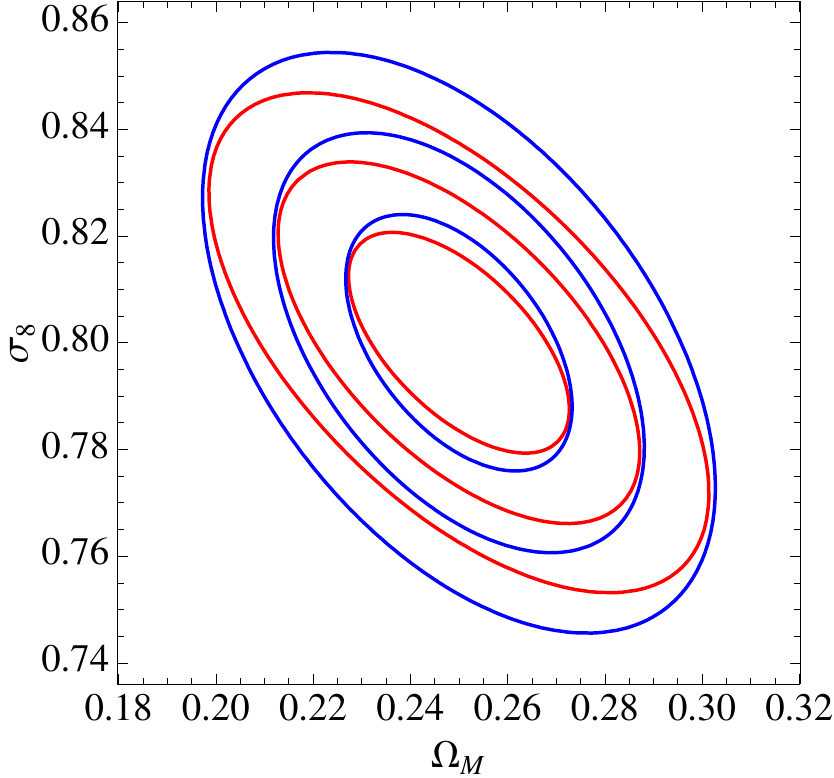}
\includegraphics[width=5.5cm]{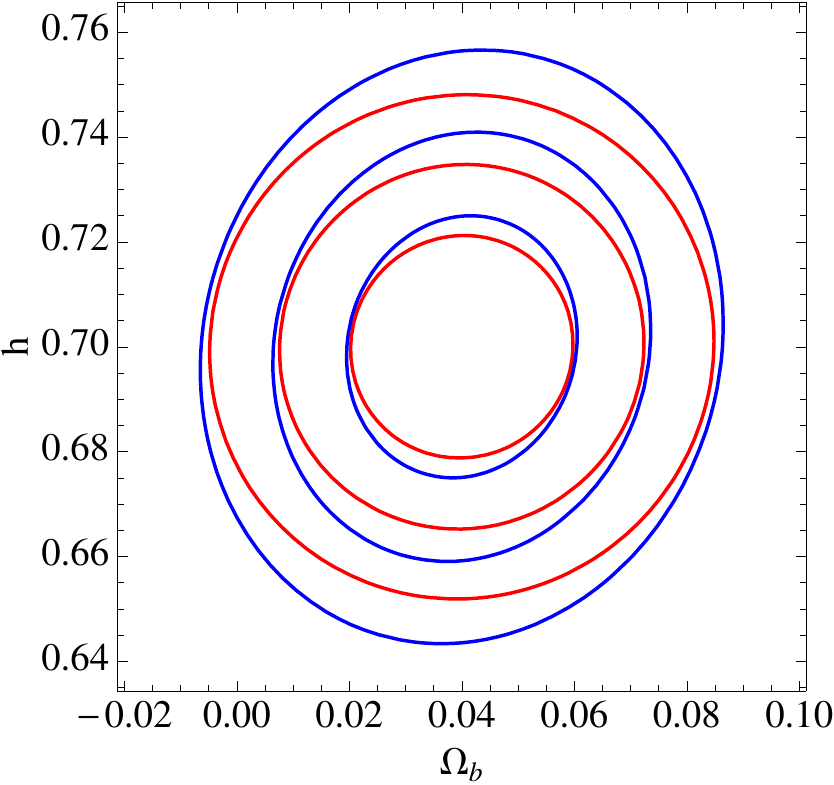}
\includegraphics[width=5.5cm]{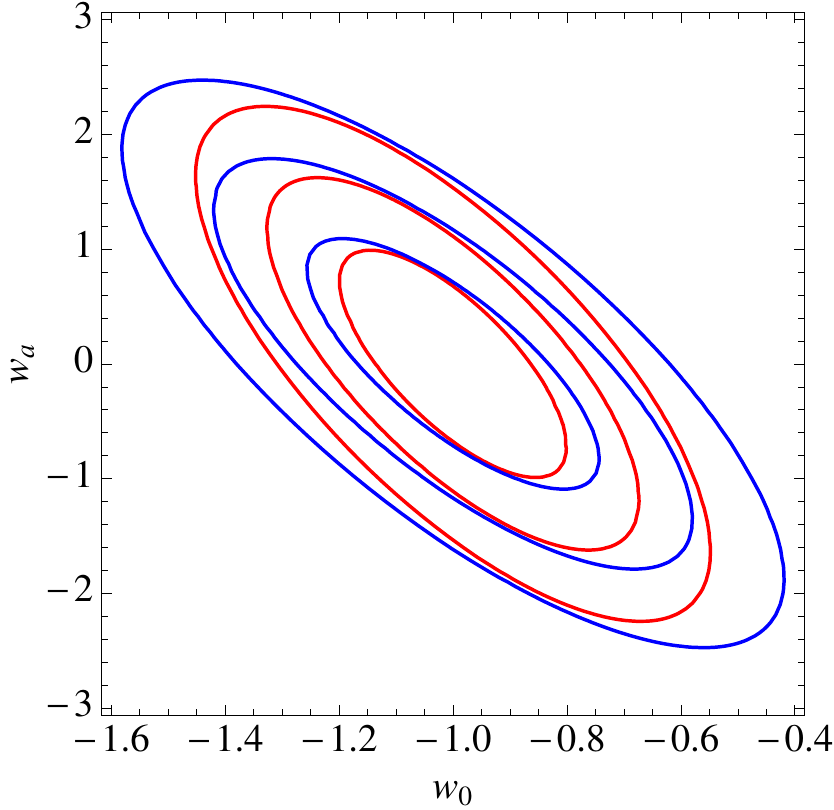}
\caption{68.3, 95.5, and $99.7\%$ confidence range in the $(\Omega_{\rm m}, \sigma_8)$, $(\Omega_{\rm b}\,-\,h)$, $(w_0, w_a)$ planes after marginalizing over the remaining parameters. Blue and red lines refer to the ST and ST\,+\,MF cases, respectively. For MFs, we use the $\vec{D}_1 \cup \vec{D}_2$ data set. \label{fig: 2Dcont}}
\end{figure*}

As an additional comment, it is worth spending some words on the counterintuitive slope of the FoM ratio vs survey area curves. While it is true that both the ST and MFs FoM increase with $\Omega$, what we are plotting is the ratio among them. As $\Omega$ gets larger, the statistical noise in the ST decreases faster than that of MFs so that the ST FoM increases faster than the MF one thus leading to a ratio which is decreasing with the area. Qualitatively, this can also be explained noting that, as the survey area increases, the reduction of the statistical uncertainties has a greater impact than the partial degeneracy breaking offered by MFs so that the ratio between the FoMs approaches a constant value which is what we indeed find.

It is worth wondering what makes the FoM increases. As an example, Fig.\,\ref{fig: 2Dcont} shows the confidence ranges in three projection of the seven dimensional cosmological parameters space after marginalizing over the other cosmological and nuisance parameters. The MFs shrink and rotate the contours with respect to the ST only case. This suggests that the increase of the FoM is the result of a chain reaction. For instance, MFs make the $(\Omega_{\rm b}, h)$ contours rounder thus partially breaking this degeneracy. As a consequence, both these parameters are better constrained which causes an improvement also of, e.g., $w_a$ since this is correlated with both $\Omega_{\rm b}$ and $h$. A stronger constraint on $w_a$ improves the contraint on $w_0$ too leading to a smaller area of the ellipses in the $(w_0, w_a)$ plane hence a larger FoM.

\section{Conclusions}

MFs have been widely used in the literature to quantify the topological properties of the CMB maps with the final aim of extracting the embedded information on the non-Gaussianity of the matter density contrast. Such a feature makes them also an attractive probe in weak lensing, since the convergence field is manifestly non-Gaussian. Todays classical shear tomography is unable to access this information so that MFs stand out as an ideal candidate to complement the analysis of the field properties and increase the overall FoM. Unfortunately, MFs are hard to compute theoretically being expressed by a perturbative expansion whose higher-order terms ask for modelling the ${\cal{N}}$-points correlation functions and their Fourier-space counterparts. While this is well known for ${\cal{N}} = 2$ and reasonably modelled for ${\cal{N}} = 3$, things get more and more untractable as ${\cal{N}} \ge 4$ with nonlinearities impossible to deal with. Moreoever, the impact of the mapmaking procedure (from the measured ellipticity to the shear field and then the convergence field) and observational effects can not be included in the theoretical approach so that one has to look for alternative techniques. 

As a possible way out, one can rely on a suite of ray-tracing N-body simulations replicating the galaxy sky positions, redshifts, and shape noise and spanning a given parameter space \cite{Petri2015}. This emulator approach allows to bypass any theoretical computation at the cost of reducing the parameter space to only three quantities, namely $(\Omega_{\rm m}, w_0, \sigma_8)$ thus cutting out most of the dark energy models of interest. Moreover, this method is survey dependent so that it can not be used to explore what happens if a different setup is adopted. 

We have therefore developed a different approach based on two main working assumptions. First, we cut the MFs expansion to first order in $\sigma_0$ so that the MFs only depend on the variance and the generalized skewness parameters of the convergence field and its derivatives. Second, splitting up the observed field in the sum of a biased reconstructed field and uncorrelated noise, we have found out analytical relations relating the theoretical MFs with the measured ones. Contrasting these relations with simulated maps including noise, has made it possible to validate the relations, thus showing that such an approach is well founded. 

Having shown that the analytical relations we have found are a reliable way to predict the MFs, we have been able to compute the Fisher matrix to investigate which constraints can be set on cosmological parameters from MFs alone and in combination with standard shear tomography considering different MFs data sets. It turns out that MFs improve the FoM by a factor of $20-30\%$ for a survey area of $\Omega=15\,000 \, {\rm deg}^2$ so that they are indeed interesting probes to improve our understanding of dark energy properties. It is worth noting, however, that we have here taken as reference an idealized survey with properties set by the available MICE simulations. As such, we have used a tomographic approach with six bins only (instead of ten as for \Euclid) and a redshift distribution which is quite flat over three contiguous redshift ranges spanning $0.1 \le z \le 1.4$ in total. This is quite different from the planned \Euclid survey which will be deeper (median redshift $z_m = 0.9$ and $0.1 \le z \le 2.0$) and with a larger number density (30 instead of $27\, {\rm {arcmin}^{-2}}$). Moving from a MICE-like to a truly Euclid-like survey will change the shear FoM so that it is worth reconsidering the present analysis. Note that this will also ask for a recalibration procedure in order to verify that our approach is still valid for different assumptions on the redshift distribution and the noise properties. 

A point that we have not addressed here is whether the results depend on the map reconstruction method adopted. Following common practice, we have here used the popular KS approach, but alternative techniques have been developed and tested on both real and simulated data \cite{Jeffrey2018}. It has been shown that the reconstructed maps are not identical and have different statistical properties. We can therefore anticipate that the topological properties and hence the MFs will be different too. Such a circumstance could also not impact our analysis since we only need the calibration relation to be valid no matter the exact values of the nuisance parameters (which are marginalized over). However, it is also possible that the noise properties are different, with each method introducing a certain degree of correlation between the noise and the signal. As far as this is negligible, we can still rely on our analytical relations provided that their validity is verified by contrasting against the convergence field reconstructed with different methods from the same simulated shear field. 

A more serious concern is the impact of masking since it introduces holes in the map, which clearly alters the topological properties and potentially biases the MFs. Although these holes are not correlated with the convergence signal, being rather related to the position of stars and/or bad pixels in the CCD, correcting them is nevertheless mandatory. The problem will then be checking whether methods such as inpainting bias the recovered MFs. We stress again, however, that we do not need to reconstruct the field, but only fill the holes in such a way that any potential bias can still be absorbed in a change of the nuisance parameters with no bias on the cosmological ones. This will be the issue of a forthcoming publication. 

While the above issues originate from observational effects, there are two further topics of astrophysical origin that need to be addressed. First, we have here neglected the intrinsic alignment (IA) which adds two contributing terms to the shear power spectrum. How IA affects the convergence bispectrum is still an open question, but we nevertheless note that the IA power spectrum is subdominant over the scale mainly contributing to the MF integral so that we do not expect a significant bias. Similarly, baryonic effects damp the matter power spectrum (and likely the bispectrum too) only at large $k$ which in the Limber approximation corresponds to large $\ell$ or small $z$. This will bias low the variance and the generalized skewness so that the overall effect on the MFs can be likely described as a multiplicative bias and reabsorbed in the formalism we have adopted. Although a detailed analysis is needed, we therefore qualitatively argue that neither IA or baryons challenge the validity of our approach. 

The results discussed in the text have to be considered as preliminary ones. Next step would be to replace the MICEv2 catalog with mock data mimicking actual ongoing or next to come surveys in order to first check whether the calibration formulae we have successfully tested are still valid with more up to date simulations. We do expect this is indeed the case having our formulae been derived in a survey independent way. Nevertheless, such a check is necessary to also get the fiducial values of the nuisance parameters to be used as input in the Fisher matrix analysis. Moreover, since we have investigated the relative increase of the FoM rather than the absolute one (i.e., we have compute the ratio of the FoM values with and without the MFs contribution), we need to know the specifics of each survey (e.g., source redshift distribution and total number density) to compute the reference FoM. This is outside the aims of this first analysis, but we will address all of these issues in a forthcoming publication.

As preliminary as they inevitably are, the results discussed in this paper can nonetheless be taken as a clear evidence that bringing topology in the realm of lensing help shedding further light on the dark energy mystery.

\begin{acknowledgments}

We acknowledge the use of data from the MICE simulations, publicly available at https://cosmohub.pic.es/home; in particular, we thank the MICE team for allowing us to use the MICECAT v2.0 catalog prior of publication. This work has made use of CosmoHub. CosmoHub has been developed by the Port d'Informaci\'{o} Cient\'{\i}fica (PIC), maintained through a collaboration of the Institut de F\'{\i}sica d'Altes Energies (IFAE) and the Centro de Investigaciones Energ\'{e}ticas, Medioambientales y Tecnol\'{o}gicas (CIEMAT), and was partially funded by the ``Plan Estatal de Investigaci\'{o}n Cient\'{\i}fica y T\'{e}cnica y de Innovaci\'{o}n" program of the Spanish government. MV and IT are funded by Funda\c{c}\~ao para a Ci\^encia e a Tecnologia (FCT) through the Investigador FCT research contract IF/01518/2014 and POPH/FSE (EC) by FEDER funding through the program COMPETE, and acknowledge additional support from FCT through research grant UID/FIS/04434/2013. VFC and XE are funded by Italian Space Agency (ASI) through contract Euclid - IC (I/031/10/0) and acknowledge financial contribution from the agreement ASI/INAF/I/023/12/0. We acknowledge the support from the grant MIUR PRIN 2015 ”Cosmology and Fundamental Physics: illuminating the Dark Universe with Euclid”. 

\end{acknowledgments}

\end{document}